\documentclass{aa}  

\usepackage{graphicx}
\usepackage{txfonts}
\usepackage{natbib}
\usepackage{tablefootnote}
\usepackage{color}

\def\e20{$\times 10^{20}$}

\def\10K{10K-sample} 
\def\20K{20K-sample} 

\def\kmsecmpc{km s$^{-1}$ Mpc$^{-1}$} 
\def\ie{i.e.,~}

\begin{document} 

   \title{An high definition view of the COSMOS Wall at $z\sim 0.73$ 
       \thanks{Based
       on observations collected at the European Southern Observatory,
       Cerro Paranal, Chile, using the Very Large Telescope under
       program ESO 085.A-0664. 
}
}
   \subtitle{}

   \author{A.~Iovino \inst{1},
          V.~Petropoulou \inst{1},
          M.~Scodeggio \inst{2},
          M.~Bolzonella \inst{3}, 
          G.~Zamorani \inst{3}, 
          S.~Bardelli \inst{3}, 
          O.~Cucciati \inst{3}, 
          L.~Pozzetti \inst{3}, 
          L.~Tasca  \inst{4}, 
          D.~Vergani \inst{5}, 
          E.~Zucca \inst{3}, 
          A.~Finoguenov \inst{6,7}, 
          O.~Ilbert \inst{4}, 
          M.~Tanaka \inst{8},
          M.~Salvato \inst{9}, 
          K.~Kova\v{c} \inst{10},   
          P. Cassata \inst{11}
          }
\offprints{Angela Iovino (angela.iovino@brera.inaf.it)}

\institute{
INAF - Osservatorio Astronomico di Brera, Via Brera, 28, I-20159 Milano, Italy \\ 
\email{angela.iovino@brera.inaf.it}
\and
{INAF - IASF Milano, Via Bassini 15, I-20133, Milano, Italy}  
\and
{INAF - Osservatorio Astronomico di Bologna, via Ranzani 1, I-40127 Bologna, Italy}  
\and 
{Laboratoire d'Astrophysique de Marseille, CNRS-Universit{\'e} d'Aix-Marseille, 38 rue F. Joliot Curie, F-13388 Marseille, France}  
\and 
{INAF - IASF Bologna, Via P. Gobetti 101. I-40129 Bologna, Italy}  
\and 
{Department of Physics, University of Helsinki, G. Hallstromin katu 2a, FI-00014 Helsinki, Finland} 
\and 
{CSST, University of Maryland, Baltimore County, 1000 Hilltop Circle, Baltimore, MD 21250, USA} 
\and 
{National Astronomical Observatory of Japan, 2-21-1 Osawa, Mitaka, Tokyo 181-8588, Japan} 
\and 
{Max Planck Institute for Extraterrestrial Physics, Giessenbachstr. 1, D-85748 Garching, Germany} 
\and 
{Institute of Astronomy, ETH Zurich, CH-8093, Z\"urich, Switzerland} 
\and 
{Instituto de Fisica y Astronom\'ia, Facultad de Ciencias, Universidad de Valpara\'iso, Playa Ancha, Valpara\'iso, Chile} 
             }

   \date{20-June-2016}

 
  \abstract
{} 
{We present a study of a large filamentary structure at z$\sim$0.73 in
  the field of the COSMOS survey, the so-called COSMOS Wall. This
  structure encompasses a comprehensive range of environments from a
  dense cluster and a number of galaxy groups to filaments, less dense
  regions, and adjacent voids. It thus provides a valuable laboratory for the
  accurate mapping of environmental effects on galaxy evolution at a
  look-back time of $\sim$6.5 Gyr, when the Universe was roughly half
  its present age.}
{We performed deep spectroscopic observations with VIMOS at VLT of a
  K-band selected sample of galaxies in this complex structure,
  building a sample of galaxies complete in galaxy stellar mass 
  down to a lower limit of log$({\cal M}_{*}/{\cal M}_{\odot})\sim
  9.8$, which is significantly deeper than previously available data. Thanks to
  its location within the COSMOS survey, each galaxy benefits from a
  wealth of ancillary information: HST-ACS data with I-band exposures
  down to $I_{AB}$$\sim$28 complemented by extensive multiwavelength
  ground- and space-based observations spanning the entire
  electromagnetic spectrum.} 
{In this paper we detail the survey strategy and weighting scheme adopted
  to account for the biases introduced by the photometric preselection of our
  targets. 
  We present our galaxy stellar mass and rest-frame magnitudes
  estimates together with a group catalog obtained with our new data and
  their member galaxies color/mass distribution.} 
{Owing to to our new sample we can perform a detailed, high definition mapping of the complex COSMOS Wall structure. The sharp
  environmental information, coupled with high quality spectroscopic
  information and rich ancillary data available in the COSMOS
  field, enables a detailed study of galaxy properties as a
  function of local environment in a redshift slice where
  environmental effects are important, and in a stellar mass range where mass
  and environment driven effects are both at work.}


   \titlerunning{ The COSMOS Wall in high definition} 
    \authorrunning{A. Iovino et al. }

   \maketitle
%

\section{Introduction}
\label{Introduction}

It is well known that the Universe as a whole formed stars more actively in
the past than today. The cosmic star formation rate (SFR) peaks at $z \sim
1$-$3$ and then decreases by an order of magnitude toward the present day
\citep{Madau1998, Lilly1996, Hopkins2006, Shioya2008, Bouwens2009,
  Cucciati2012, Madau2014}. The evolution of the cosmic SFR is an important
observational constraint for all the current models of galaxy formation and
evolution \citep{Somerville2012}.

Among the different mechanisms that regulate SFR evolution, galaxy
  stellar mass is a well-known key factor: galaxies of different
stellar masses evolve on different timescales according to the
so-called downsizing trend \citep{Cowie1996, Kauffmann2004}.

In the local Universe star formation (SF) activity in galaxies also
depends on environment. High density regions, such as groups and
clusters of galaxies, are dominated by passively evolving early-type
galaxies, while there are many star-forming late-type galaxies in
field environments \citep{Dressler1980, Goto2003, Bamford2009}. Thus,
the fraction of star-forming galaxies systematically decreases with
increasing local galaxy density \citep{ Gomez2003, Balogh2004B,
  Kauffmann2004, Tanaka2004}. This effect is also observed when
splitting the analysis into stellar mass bins \citep{Baldry2006} and
thus removing the effect due to high-mass galaxies segregating in the
densest environments \citep{Scodeggio2009, Bolzonella2010}.  It is
still a matter of debate how the relation between SF and density
evolves with cosmic time and what mechanisms drive this evolution
(secular as opposed to nurture driven evolution).

 By focusing on star-forming galaxies, it is generally found that
  environment does not significantly affect the SFR-stellar mass relation
  \citep{Peng2010, Koyama2013, Brough2013} and this fact, coupled with the
  environmental effect on the fraction of star-forming galaxies, has been used
  to advocate a rapid star formation quenching mechanism to explain a galaxy
  transition from star forming to passive \citep{Brough2013}.  On the other
  hand, by examining the global galaxy population it has been shown that,
  since $z \sim 1$, galaxies in groups have been converted more rapidly than
  the global field population from the blue, star-forming, late-type morphology
  cloud to the red, quiescent, early-type morphology sequence \citep{Iovino2010,
    Kovac2010, McGee2011, Peng2012, Popesso2015, Guglielmo2015}.  In
  particular, the \citet{Popesso2015} modeling of this trend  seems to favor a
  relatively slow star formation quenching mechanism that takes place on
  timescales of 1 Gyr or more, while \citet{Guglielmo2015} advocate an
  accelerated star formation in high-mass haloes to explain the trend.

  It has also been argued that the SF-density relation should reverse
  once we reach the epoch when early-type galaxies formed the bulk of
  their stars ($z\gtrsim 1.5$). During that phase high density regions
  should host highly star-forming galaxies, but the observational
  evidence for this is still debated
  \citep[see][]{Ziparo2014,Mei2015}.  Thus, defining the role played
  by environmental processes and pinning down which physical
  mechanisms play an important evolutionary role in dense regions is
  not a simple task.

The search for the transition population, that is the galaxies with
colors that are intermediate between the blue cloud of star-forming
galaxies and the passive sequence of galaxies with negligible present
star formation, has been particularly elusive.  The search for an
environmental dependence on the frequency of such population
\citep{Mok2013, Erfanianfar2015}, a dependence that could provide
insights into the timescale of the mechanisms at work and useful
constraints for simulations, has been equally elusive.

The Cosmological Evolution Survey (COSMOS) offers a unique opportunity
to explore this topic: it contains a very complex structure located at
$z \sim 0.73$ and so conspicuous that it was clearly evident in the
COSMOS photometric catalog \citep{Scoville2007c}. In an extremely
narrow redshift slice, $0.72\leq z_{spec}\leq 0.74$, there is a rich
X-ray detected cluster and a number of groups, some of which are X-ray
emitting. These groups are embedded in a filamentary structure
extending across $\sim 20$ Mpc (comoving) within the survey field, and
are surrounded by significantly lower density foreground and
background regions \citep[see][]{George2011}.

Its location at a look-back time of $\sim 6.5$ Gyr is at a crucial
transition epoch when the color-density relation becomes detectable
for galaxies of mass log$({\cal M}_{*}/{\cal M}_{\odot})\sim 10.5$,
while the morphology-density relation remains undetected
\citep{Iovino2010, Kovac2010}.

Similar depth observations at higher redshifts cannot serve our goals,
as at $z\sim1$ only massive galaxies (log$({\cal M}_{*}/{\cal
  M}_{\odot})\ga 11.0$) are detectable in current redshift survey
samples. These galaxies are quite rare because of the exponential
cutoff at the high-mass end of the galaxy stellar mass function
\citep{Ilbert2010}; they are also more rapidly evolving
\citep{Noeske2007b, Thomas2010,Gilbank2011}. Their intrinsic,
mass-dependent evolutionary timescale is shorter than the
environmental effects. As a result, red galaxies dominate the galaxy
stellar mass function in all environments.

In this paper we present the results of an observational campaign
targeting this interesting structure. The goal was to obtain a sample
complete in galaxy stellar mass down to log$({\cal M}_{*}/{\cal
  M}_{\odot})\sim 9.8$ within this region of the Universe, doubling
the numbers of galaxies already available from previous spectroscopic
follow-up in this area \citep{Lilly2007}.

These new spectroscopic data enables us both to extend previous
analysis to yet unexplored galaxy stellar mass ranges and to define
the environment much more sharply. The new sample we present therefore
offers the opportunity for a detailed study of galaxy properties as a
function of local environment in a redshift slice where environmental
effects have been shown to be important, and in a mass range where
mass and environment driven effects are both at work.

A fiducial $\Lambda$ cold dark matter cosmology model is assumed
throughout our paper with $H_0 = 70$ \kmsecmpc, $\Omega_{m} =
0.25,$ and $\Omega_{\Lambda } = 0.75$.  All magnitudes are always
quoted in the AB system \citep{Oke1974}.

\section{The COSMOS Wall at $z\sim 0.73$} 
\label{Walldescription}

COSMOS is a panchromatic imaging and spectroscopic survey of a
$1.4\times1.4$ ~deg$^2$ field, which is designed to probe galaxy
formation and evolution as a function of cosmic time and large-scale
structure environment \citep{Scoville2007a}.  The core of the COSMOS
survey consists of an HST Treasury Project \citep{Scoville2007b}, but
the equatorial location of the COSMOS field has offered the critical
advantage of allowing major observatories from both hemispheres to
observe this field.

At the very beginning of the COSMOS project, photometric data alone
were sufficient to detect a prominent large-scale structure located at
$z\sim 0.7$ \citep{Scoville2007c,Cassata2007, Guzzo2007}. Nevertheless, the
low precision of photometric redshifts limited the detailed
exploration of the dependency of local environment on quantities, such as
galaxy color or galaxy morphology, with the result that galaxy mass
appeared to be the main, if not unique, driver of the observed
trends.

Since those initial studies, the ancillary photometric and
spectroscopic data available in the COSMOS field have increased
significantly. The 30-COSMOS photometric reference catalog, now
contains magnitudes measured in 30 bands. These include 2 bands from
the Galaxy Evolution Explorer (GALEX), 6 broadbands from the
SuprimeCam/Subaru camera, 2 broadbands from MEGACAM at CFHT, 14 medium
and narrowbands from SuprimeCam/Subaru, J band from the WFCAM/UKIRT
camera, K band from the WIRCAM/CFHT camera, and the 4 IRAC/Spitzer
channels. These magnitudes were obtained using PSF-matched photometry
for all the bands from the u to the K band, measured over an aperture
of $3\arcsec$ diameter at the position of the Subaru i+ band
detection, while the corresponding magnitudes in the IRAC bands
($3.6$, $4.5$, $5.6,$ and $8.0\mu m$) were obtained following the
procedure described in detail in \citet{Ilbert2009}. The imaging data
are extremely deep, reaching $u \sim27$, $i+\sim26.2$, and
$K_{s}\sim23.7$ for a 5$\sigma$ detection in a 3'' diameter aperture
\citep[the sensitivities in all bands can be found in][]{Capak2007,
  Salvato2009}. Highly accuracy photometric redshifts based on this
photometric data set were derived using an improved version of the
software {\it LePhare}, which accounts for the contributions from
emission lines (see www.oamp.fr/people/arnouts/LE\_PHARE.html). The
error in the photometric redshift estimate is as low as $0.007/0.012
\times(1+z) $ for galaxies brighter/fainter than $I_{AB} = 22.5,$
respectively \citep{Ilbert2009}. The 30-COSMOS photometric catalog
contains 1,500,515 sources in total, and 937,013 sources at $i+<26.5$
\citep{Ilbert2009}.

The COSMOS field has also been observed in the far-infrared with the
Multiband Photometer for Spitzer (MIPS) at $24$, $70$, and $160 \mu m$
\citep{Sanders2007}; in the radio continuum with deep 1.4 GHz and 324
MHz observations as part of the VLA-COSMOS survey
\citep{Schinnerer2007, Smolcic2014}; and in the X-ray with both the
XMM-Newton and the Chandra satellites \citep{Hasinger2007, Elvis2009},
while deep UltraVISTA observations are ongoing \citep{McCracken2012}.

A number of spectroscopic surveys have targeted this field.  Publicly
available data are provided by the PRIMUS survey ($\sim 30,000$
redshifts in the COSMOS field), but with redshift measurement errors
that are too large to allow a precise definition of the local
environment \citep{Coil2011}.  A further public set of redshift
measurements is that by \citet{Comparat2015} and was obtained with the
FORS2 spectrograph at the ESO Very Large Telescope and targeted color
selected emission line galaxies ($\sim\,2000$ redshift
measurements). Finally, the zCOSMOS Bright survey \citep{Lilly2007},
which was carried out with the VIMOS spectrograph at the ESO Very
Large Telescope, covers the whole area of the COSMOS field and
provides redshifts for $\sim\,20000$ galaxies. These galaxies were
selected from a purely magnitude-limited sample down to
$I_{AB}\,\leq\,22.5$, as measured from the HST-ACS imaging, with a
velocity uncertainty of $\sim$\,$100$\,km\,s$^{-1}$. This last has
been designated as the \20K \citep{Lilly2009}, and the group catalog
derived using these galaxies lists six groups with ten or more
spectroscopically confirmed members inside the large-scale structure
located at $z\sim 0.73$, thus confirming early photometric results
\citep{Knobel2012}.

    \begin{figure}
    \centering
    \vspace{-1.6cm}
    \includegraphics[width=7.5cm, angle=0]{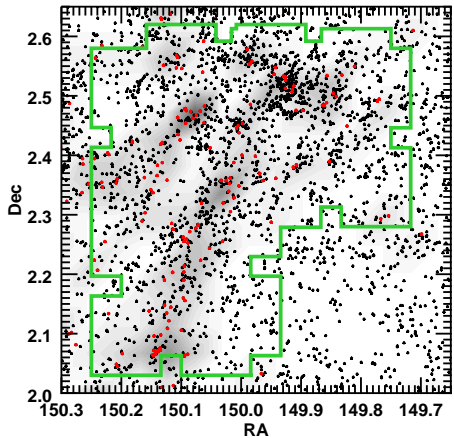}
    \vspace{-1.5cm}
    \caption{RA-Dec distribution of galaxies within the area we targeted by
      VIMOS observations, indicated by the green outline. Black dots represent
      galaxies of the 30-COSMOS photometric catalog with $0.60\leq
      z_{phot}\leq 0.86$, while red dots (and the underlying gray contours)
      represent galaxies from the \20K in the range $0.72 \leq z_{spec} \leq
      0.74$. It is possible to see the presence of a significant galaxy
      concentration and of two long filamentary structures pointing toward
      it.}  \label{Fig1}
     \end{figure}

In Figure~\ref{Fig1} black dots show the distribution on the sky of
the 30-COSMOS galaxies brighter than $K_{s} = 22.6$ and with
photometric redshift $0.60 \leq z_{phot} \leq 0.86$. Despite the broad
photometric redshift range chosen, a clear concentration of galaxies
is visible at the top center of the field, together with two
higher density filaments starting from it and extending toward the
southeast. The density contrast of the filamentary structures is more
striking when considering only galaxies from the \20K with
spectroscopic redshift in the range $0.72 \leq z_{spec} \leq 0.74$,
shown as red dots and gray contours in Figure~\ref{Fig1}.

\begin{figure}
    \centering
    \vspace{0cm}
    \includegraphics[width=7.0cm, angle=270]{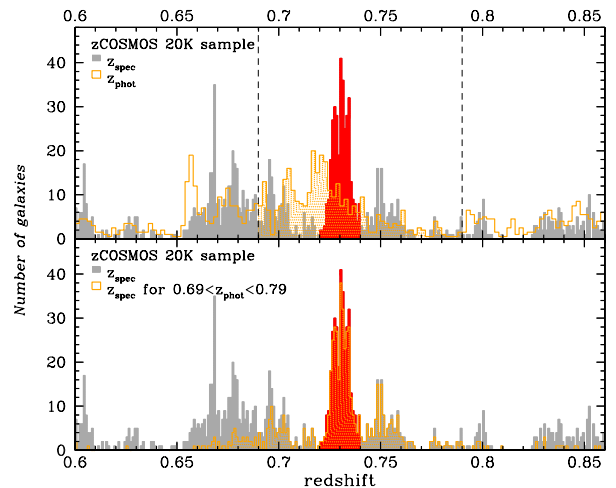}
    \caption{Top panel: redshift histogram of the \20K within the sky
      region targeted for observations. In red we highlight the subset
      of galaxies in the redshift range $0.72 \leq z_{spec} \leq
      0.74$, corresponding to the prominent peak of the COSMOS Wall
      structure. For comparison the orange histogram shows the
      photometric redshift distribution of the full \20K within the
      same sky region. The Wall volume redshift interval, $0.69 \leq z
      \leq 0.79$ (indicated by two dashed lines), includes both the
      COSMOS Wall structure and two conspicuous lower density regions
      in the redshift ranges $0.71 \leq z_{spec} \leq 0.72$ and $0.76
      \leq z_{spec}\leq 0.77$.
      \newline Bottom panel: same as top panel but with superimposed (in
      orange) the histogram of the spectroscopic redshift distribution of the
      galaxies possessing a photometric redshift within the Wall volume:
      $0.69\leq z_{phot}\leq 0.79$.} \label{Fig2}
     \end{figure}

In the top panel of Figure~\ref{Fig2} the filled gray histogram shows
the spectroscopic redshift distribution of the \20K galaxies within
the sky region highlighted in Figure~\ref{Fig1}. In red we highlighted
the subset of galaxies in the redshift range $0.72 \leq z_{spec} \leq
0.74$, corresponding to the prominent peak of the COSMOS Wall
structure.
Adjacent to this high density redshift slice there are two emptier regions in
the redshift ranges $0.71 \leq z_{spec} \leq 0.72$ and $0.76 \leq z_{spec}
\leq 0.77$, as shown by the dip in numbers of the histogram of
Figure~\ref{Fig2}. 

From now on we call Wall volume the region of volume in space defined in
RA-Dec by the polygonal boundaries on the sky shown in Figure~\ref{Fig1},
that is the outline of the different VIMOS pointings of our observations (see
Section~\ref{ObsStrategy}), and defined in redshift by the limits $0.69 \leq z \leq
0.79$.  This redshift interval was somewhat arbitrarily chosen to
  encompass the largest and most prominent groups in the region, and the
  filaments that connect them, plus the two lower density regions nearby.  In
  this way we can optimally sample a large range of local galaxy densities
  within a redshift range narrow enough to allow us to neglect evolutionary
  effects as a function of cosmic time.

In the final \20K, there are 658 galaxies possessing a reliable spectroscopic
redshift within the Wall volume, out of which 350 are within the peak of what
we  call from now onward the Wall structure, located in the narrow
redshift range $0.72 \leq z_{spec} \leq 0.74$ (highlighted in red in
Figure~\ref{Fig2}). The \citet{Comparat2015} sample contains only 13
  additional galaxies within the Wall volume, so, given the complexity of
  incorporating its different selection criterion in our analysis, we choose
  to neglect these measurements in the following.

\section{The enlarged COSMOS Wall sample}  
\label{targetselection}

We targeted the COSMOS Wall structure with a spectroscopic observational
campaign using VIMOS.  In the following sections, we describe in detail the
magnitude and photometric redshift preselection strategy adopted to
optimally define the parent catalog of our observations.

\subsection{K-band target selection}  
\label{massselection} 

An efficient way to select a galaxy sample complete in stellar mass up
to $z \sim 1.0$ is to use near-infrared (NIR) $K$-band photometry. The galaxy
emission observed in that band is dominated by the old stellar
population that represents the majority of the total stellar mass
budget, rather than by the products of the recent star formation
activity.

The 30-COSMOS catalog \citep{Ilbert2009} is providing WIRCAM $K_s$
band aperture magnitudes for galaxies brighter than $K_s \sim 24.0$,
together with a rough estimate of the galaxy stellar mass, obtained
using the analytical relations from \citet{Arnouts2007}. To improve
the quality of the stellar mass estimates  and for homogeneity
  with our analysis later on, we decided to recompute these estimates with a
full spectral energy distribution (SED) fitting technique, based on an updated version of Hyperzmass
\citep{Bolzonella2000, Bolzonella2010}.  The SED templates adopted in
this procedure are derived from simple stellar populations (SSPs)
modeled by \citet{BruzualCharlot2003}, adopting the
\citet{Chabrier2003} initial mass function \citep[IMF; see for more
  details][]{Bolzonella2010}.

\begin{figure}
    \centering
    \vspace{-2.5cm}
    \includegraphics[width=7.5cm, angle=0]{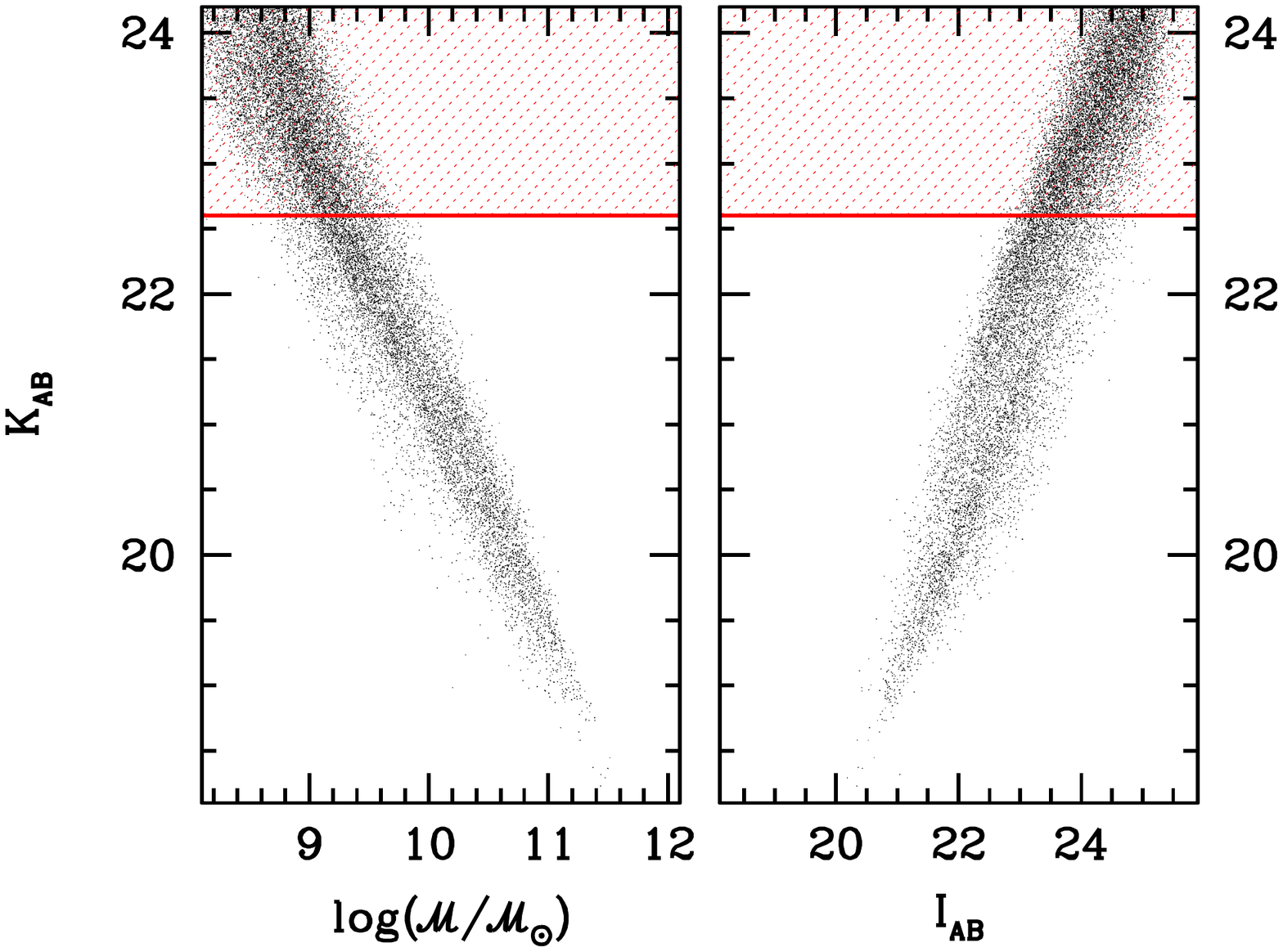}
    \vspace{-2cm}
    \caption{Both Panels show galaxies from the COSMOS photometric
      catalog with $0.69\leq z_{phot}\leq 0.79$.  The left Panel shows
      $K_s$-band WIRCAM magnitudes versus masses (computed
      using Chabrier IMF, and Bruzual\&Charlot model library, no
      secondary bursts). The horizontal red line shows our magnitude
      cut-off $K_s \leq 22.6$, that enables us to reach completeness,
      irrespective of colors, for masses down to the limit log$({\cal
        M}_{*}/{\cal M}_{\odot})= 9.8$.  The right Panel shows 
      $K_s$-band WIRCAM magnitudes versus $I_{AB}$ magnitudes.  Within
      our $K_s$-band selection limit the vast majority of galaxies are
      brighter than $I_{AB} = 24.0$.}
               \label{Fig3} 
     \end{figure}

We then used these stellar mass estimates to check how a
  $K_s$-band preselection translates into a galaxy stellar mass
selection in the redshift range of interest $0.69\leq z_{phot}\leq
0.79$.  The left panel of Figure~\ref{Fig3} shows that a selection
down to $K_s \leq 22.6$ is sufficient to remain complete, 
irrespective of colors, for galaxy stellar masses down to the limit
log$({\cal M}_{*}/{\cal M}_{\odot})= 9.8$.

The right panel of Figure~\ref{Fig3} shows that the
percentage of galaxies at $K_s \leq 22.6$ that are within the photometric
redshift range $0.69 \leq z_{phot}\leq 0.79 $  and are fainter than
$I_{AB} = 24.5$ (well above the 5$\sigma$ detection limit of the $i+$
Subaru observations), is less than $0.6\%$ and thus negligible. We can
therefore be confident that the chosen $K_s$-band selection limit
($K_s = 22.6 $) is well within the range of VIMOS spectrograph
capabilities.

\subsection{Photometric redshift selection}  
\label{photzselection} 

An important ingredient for our target selection is the use of
photometric redshifts to maximize our success rate for galaxies
within the Wall volume and minimize observations of interlopers, \ie
galaxies that are outside our redshift range of interest.

The quality of photometric redshifts in the reference 30-COSMOS
photometric catalog is very good. Their error is as low as $\sigma
\sim 0.007(0.012)\times(1+z)$ down to $I_{AB} = 22.5(24.0),$
respectively \citep[see][]{Ilbert2009}, and we can optimize our target
galaxy selection by taking advantage of the available zCOSMOS \20K
spectroscopic redshifts.

A comparison between the capability of the spectroscopic and
photometric redshifts in identifying the COSMOS Wall structure is
presented the top panel of Figure~\ref{Fig2}, where we compare the
spectroscopic (in gray and red) and photometric (in orange) redshift
distribution of the galaxies in the \20K located within the RA-Dec
outline of the Wall volume (green contours in Figure~\ref{Fig1}). The
contrast between the Wall structure (red histogram) and the
foreground/background distribution is sharper when we use
spectroscopic redshifts, although the total number of galaxies
selected in the Wall volume changes only slightly when using
photometric or spectroscopic redshift values. Intuitively, as many
galaxies enter the Wall volume range as many leave owing to
uncertainties in photometric redshifts because the main peak of our
structure is located far from the selection window limits.

Because of this dilution of contrast it is important to understand how
the presence of photometric redshift uncertainties, albeit small,
may deteriorate our ability to recover the true spectroscopic redshift
distribution within this region when imposing a photometric redshift
preselection to the sample targeted for spectroscopic observations.

In the bottom panel of Figure~\ref{Fig2}, we overplot on the total
spectroscopic redshift distribution already shown in the top Panel,
the spectroscopic redshift distribution (orange shaded histogram) of
the subset of galaxies of the \20K (645 in total) possessing a
photometric redshift in the range $0.69\leq z_{phot}\leq 0.79$, \ie
within the limits of the Wall volume according their photometric
redshift.

It is apparent from this Figure that selecting galaxies in the
photometric redshift range $0.69\leq z_{phot}\leq 0.79$ enables us to
retrieve $\sim 90\%$ of the \20K galaxies within the Wall structure
(319 galaxies recovered out of 350), and to retrieve $\sim 84\%$ of
the full spectroscopic sample within the whole Wall volume (552
galaxies recovered out of 658).

At the same time, only 93 out of the 645, which are selected to have photometric redshift
in the range $0.69\leq z_{phot}\leq 0.79,$ are located outside the
spectroscopic range $0.69\leq z_{spec}\leq 0.79$, corresponding to an
approximate failure rate of the photometric redshift selection in retrieving
the correct spectroscopic redshift sample of $\sim15\%$. Obviously, as
visible from the bottom panel of Figure~\ref{Fig2}, the success rate of the
photometric redshift selection has a strong dependence on redshift within our
redshift interval of interest. The efficiency in retrieving galaxies located
roughly at the center of our photometric selection window is more than $90\%$,
while near the edges of the photometric selection window the efficiency drops to lower
values.

We can conclude that a photometric redshift preselection limiting the targets
to the photometric redshift range of the Wall volume is efficient in selecting
galaxies truly located within the Wall volume, at least for targets
brighter than $I_{AB} = 22.5$.  We need to model the loss of efficiency
in retrieving the correct redshift selection for targets fainter than $I_{AB}
= 22.5$. We discuss in detail in Section~\ref{weightingscheme} how to
parametrize what we call the photometric redshift efficiency, that is our
ability to retrieve the underlying true spectroscopic redshift distribution
at all the magnitudes of interest.

\subsection{Parent sample definition}  
\label{parentsample} 

The parent sample for our observing program (the Wall parent sample;
$WPS$) is then defined by the following conditions:
\\

\indent \indent 
      $K_{s}\leq 22.6$ \\
and \\
\indent \indent  
$0.69\leq z_{phot}\leq 0.79$ ~~or ~~$0.69\leq z_{spec}\leq 0.79$. \\         

It is therefore the full $K_{s} \leq 22.6$ WIRCAM magnitude limited
sample located within the Wall volume, either thanks to spectroscopic
redshifts available in the \20K or to COSMOS photometric
redshift information.

    \begin{figure}
    \centering
    \vspace{-2cm}
    \includegraphics[width=7.5cm, angle=0]{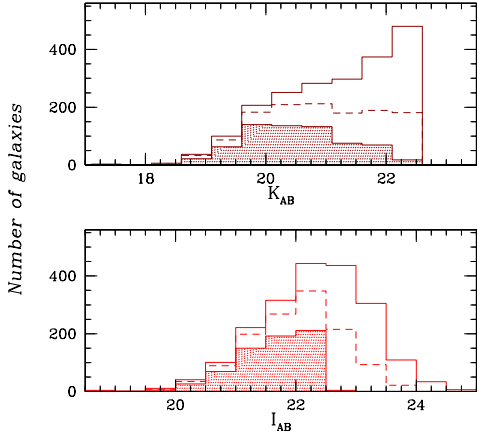}
    \vspace{-2cm}
    \caption{Top Panel: $K_s$-band magnitude distribution of the Wall parent
      sample. Bottom Panel: $I$-band magnitude distribution of the Wall parent
      sample. In both panels the shaded histogram refers to the corresponding
      magnitude distribution of the subset of galaxies possessing a reliable
      spectroscopic redshift in the \20K, while the dashed line to that 
      of our final new Wall volume sample (1277 galaxies).}
      \label{Fig4}
     \end{figure}

Inside the 30-COSMOS catalog there are 2039 galaxies satisfying these
constraints, and 658 of these already possess a reliable spectroscopic
redshift in the \20K. In the following we identify the galaxies
only possessing a photometric redshift as the $WPS_{phot}$ sample, while the
galaxies in the \20K are identified as the $WPS_{spec}$ sample.

Figure~\ref{Fig4} shows the apparent magnitude distribution of the total Wall
parent sample in the $K_s$ (top panel) and in the $I$ (bottom panel) band. The
shaded histograms in both panels refer to the 658 galaxies already possessing
a reliable spectroscopic redshift in the \20K, i.e., those whose location in
the Wall volume is already reliably assessed, while the dashed line histogram
shows the full set of 1277 galaxies that constitute out new Wall volume
sample.

\section{VIMOS observations}
\label{VIMOSobservations}

Our observing program was awarded 41 hours at the VIMOS multiobject
spectrograph, mounted on the Nasmyth focus B of ESO VLT-UT3 Melipal
(PI A. IOVINO - ESO 085.A-0664). The VIMOS imaging
spectrograph has four channels and each channel (quadrant) covers $7 \arcmin \times8
\arcmin$ with a gap between each quadrant of $\sim 2 \arcmin$. Each
quadrant is a complete spectrograph with the possibility to insert
slit masks,  as well as broadband filters
or grisms, at the entrance focal plane \citep[see][]{LeFevre2003}.

The pixel scale on the $2k\times4k$ CCD detectors is $0.205 \arcsec$
pixel$^{-1}$, providing excellent sampling of the typical image
quality at Paranal.
The observations were performed during service mode runs in 2010 and
2011. The MR grism and the OS red filter were used, together with $1.0
\arcsec$ width slits, to produce spectra with a dispersion of 2.5
$\AA$ pixel$^{-1}$ covering the spectral range 5550-9450 $\AA$ with a
spectral resolution R $\sim$ 600.

All masks were observed with the slits oriented E-W, thus different from the
original N-S orientation adopted by the zCOSMOS survey. This choice aimed at
increasing the homogeneity of the final ra-dec coverage, as the existing
$WPS_{spec}$ displayed a slight tendency to be preferentially distributed
along N-S stripes \citep[see][]{Lilly2009}..

All pointings, with the only exception of P1 and P2 (see below), were
observed after the VIMOS refurbishing that took place in the Summer
2010 (see below for pointings definition). During the refurbishing the
original thinned back-illuminated e2v detectors were replaced by
twice as thick e2v detectors, considerably lowering the fringing and
increasing the global instrument efficiency by up to a factor 2 for
wavelengths longer than $8000\AA$ \citep{Hammersley2010}. As a
consequence, the quality of the spectra obtained with the refurbished
VIMOS was significantly better.

The total area covered by our survey amounts to $0.23~deg^2$, and the
total Wall volume spans a typical transverse dimension of $\sim 21/23$
Mpc at $ z \sim 0.69/0.79$ and $\sim 300$ Mpc along the line of sight,
totalling a volume of $\sim 140\times 10^3$ Mpc$^3$.


\subsection{Pointing definition strategy}
\label{ObsStrategy}

\begin{figure} 
    \centering
    \vspace{-1.6cm}
    \includegraphics[width=7.5cm, angle=0]{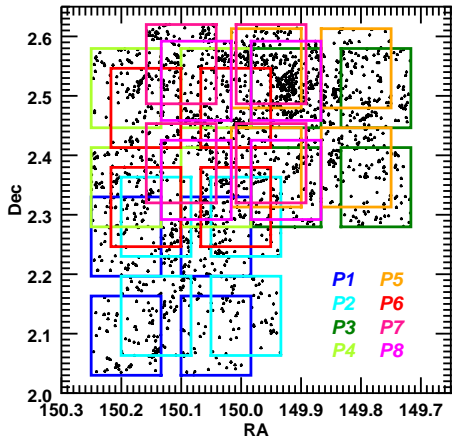}
    \vspace{-1.8cm}
    \caption{Ra--Dec distribution of the 8 VIMOS pointins (P1, P2, ..., P8)
      of our program.  Black dots represent all galaxies of
      the Wall parent catalog.  The large prominent structure on the
      top north of the field was covered by four pointings.}
       \label{Fig5}
     \end{figure}

Eight different VIMOS pointings (named P1 to P8) were used to cover the Wall
area. In Figure~\ref{Fig5} we show the position of each pointing, together
with the position of the galaxies in our parent catalog.  We tried to maximize
the number of pointings covering the denser regions on the sky while extending
the area covered as to include the filamentary distribution of galaxies
that defines the COSMOS Wall structure.
 
The range in $I_{AB}$ magnitudes covered by potential targets in our
parent catalog is very wide (see Figure~\ref{Fig4}) and we decided to
adopt a nonstandard VIMOS observing strategy to obtain spectra with
roughly the same quality for all galaxies, while maximizing the number
of targets observed.

To obtain spectra with signal-to-noise ($S/N$) values $\gtrsim $ 4 per
resolution element we decided to observe the targets brighter than
$I_{AB}= 22.5$ for at least roughly one hour, the targets in the
magnitude range $22.5 < I_{AB} \leq 23.0$ for roughly two hours, and
targets fainter than $I_{AB}= 23.0$ for more than three hours. To
achieve this we split the available $3h~37.5m$ of observing time per
each pointing into three modules (or passes) of $1h~12.5m$ each, so
that different galaxies could be observed with different integration
times depending on their $I_{AB}$ magnitude.

The number of targets for the various subsets so defined are listed in
Table~\ref{tab:Tab1}. A handful of galaxies in $WPS_{spec}$ that are fainter
than $I_{AB}= 22.5$ are secondary targets within \20K, \ie targets
observed by chance in the slit of a primary target and whose redshift
was successfully measured.

\begin{table}
\caption{Summary of Wall parent sample}
\label{tab:Tab1}  
\centering                  
\begin{tabular}{c c c c c }    
\hline\hline                 
               & $N_{tot}$ & $WPS_{spec}$ & $WPS_{phot}$ \\   
\hline \hline                    
All magnitudes            &   2039    &  658    &   1381  \\      
$I_{AB} \leq 22.5$         &   1143    &   654    &   489  \\      
$22.5 <I_{AB} \leq 23.0$   &    436    &    4    &   432  \\      
$I_{AB} > 23.0$            &    460    &    0    &   460  \\      
\hline                             
\end{tabular}
\end{table}

Our MOS slit positioning strategy was organized into two
steps. Briefly, in the first step we maximized the number of slits
positioned on the targets of the $WPS_{phot}$, \ie the subset of the
Wall parent sample possessing only photometric redshift information.
In the second step we added targets from the $WPS_{spec}$ or from the
pool of already targeted objects from $WPS_{phot}$ as fillers to our
observations, whenever possible, thus increasing their $S/N$. 

The first slit positioning step consisted in allocating in the first
pass of each pointing ($1h~12.5m$ exposure time) the maximal number of
slits that SPOC, the VIMOS slit positioning software
\citep{Bottini2005}, can allocate, starting from the set of objects in
the $WPS_{phot}$ sample.

A judicious use of the flags allowed by SPOC ({\bf C} for compulsory, {\bf F}
for forbidden, and {\bf S} for observable; see \citet{Bottini2005}), then
enabled us to force the fainter objects observed in the first pass to be
observed in the subsequent passes of each pointing, while new bright objects
could be used to replace those already observed in the previous
passes. Galaxies with $22.5<I_{AB}<23.0$ were observed in at least two
passes of $1h~12.5m$, totaling $2h~25m$ of exposure time, while galaxies with
$I_{AB}>23.0$ were observed in all the three passes of the pointing
considered, acquiring $3h~37.5m$ of exposure time.

Once this first step of our observing strategy was consolidated and the
list of targets observed in each pass for each pointing is defined
according to the rules listed above, we moved to the second phase of
our slit positioning strategy.  Here we added to the set of slit
targets, defined in the first phase, all possible extra slits that
could be positioned either on the galaxies in the $WPS_{spec}$ or on
any of the galaxies already targeted in the first step of our slit
positioning, thus appreciably increasing  the $S/N$ of their spectra.

\begin{table*}
\caption{Summary of the observed sample. The number in parenthesis
  always indicates the subset of galaxies from $WPS_{spec}$. These are
  targets that are used as fillers in our slit positioning strategy,
  or, when secondary targets, objects that just enter the slit of a
  primary target by chance.  }
\label{tab:Tab2}  
\centering                  
\begin{tabular}{l c c c c  }    
\hline\hline                 
                     &        & Primary &  &    Secondary   \\   
\hline 
                     & Observed objects   & Successful z & Inside Wall volume &  Inside Wall volume  \\   
\hline                   
All magnitudes           & 1016 (224) & 975 (224) & 836 (224) &  20 (13)  \\      
$I_{AB} \leq 22.5$        & 567 (222)  & 555 (222) & 511 (222) &  16 (12)  \\            
$22.5 <I_{AB} \leq 23.0$  & 276 (2)    & 264 (2)   & 211 (2)   &   3 (1)   \\      
$I_{AB} > 23.0$           & 173 (0)    & 156 (0)   & 114 (0)   &   1 (0)   \\      
\hline                             
\end{tabular}
\end{table*}

At the end of this procedure, we were able to position $\sim$\,$130$
slits per each pointing on average from the parent catalog.  The exact
number varies from 150 to 90, depending on the surface target
density. The total number of independent targets observed is 1016, all
providing extracted spectra for redshift measurements. We positioned
792 slits on targets from $WPS_{phot}$, while the remaining 224 were
positioned on targets from $WPS_{spec}$, which allows us to
significantly improve their spectral quality and provides a set of
galaxies useful for defining our accuracy in velocity measurements.
Our exposure time for brighter targets is $\sim 20\%$ longer than that
of the \20K observations and the new e2v CCDs, producing significantly
improved spectra, were used for most of our observations. We also
extract 120 spectra for the secondary targets, which are targets
observed by chance in a slit centered on another object. We expect,
however, that only a small percentage of this set of objects to be
within the Wall volume, as they represent a random sampling of the
global galaxy population.


\section{Data reduction and redshift measurements}
\label{data_reduction}

Data reduction for each mask was carried out using the VIPGI
software \citep{Scodeggio2005}.

After each CCD frame is bias subtracted, the precise location of the
two-dimensional (2D) spectrum for each slit on each frame of the
jitter sequence is measured and the position of individual object
spectra inside the slit is recorded. The sky background subtraction is
carried out for each slit independently, using a two-step procedure.
First, the median background level at each wavelength is measured in a
single exposure and subtracted from the data; then the median of the
sky subtracted spectra is obtained without applying any correction for
the telescope pointing offset between exposures. This median produces
a frame from which the spectra of individual objects are eliminated,
but that includes all residuals not corrected by the first sky
subtraction step, and in particular the fringing pattern, which varies
with position across the slit and wavelength and is then subtracted
from the individual 2D spectra.

The wavelength calibration is obtained via the observation of helium
and argon arc lamps through the observed masks, carried out
immediately after the sequence of science exposures. Wavelength
calibration spectra are extracted at the same location as the object
spectra and the lamp lines are identified to derive the pixel to
wavelength mapping for each slit. The wavelength to detector pixel
transformation is fit using a fourth order polynomial, resulting in a
mean $rms$ deviation of $\sim 0.4 \AA$ across the full wavelength
range covered by the observations. Finally the sky and fringing
residual subtracted and wavelength calibrated 2D spectra are combined
together, removing the spectra position offset induced by the
jittering pattern, to derive a final combined 2D spectrum for each
slit; one-dimensional (1D) spectra are extracted from the combined 2D
spectrum, using an optimal extraction procedure, based on the object
profile \citep{Horne1986}.  Each 1D spectrum is then flux calibrated
using the ADU to absolute flux transformation computed from the
observations of spectrophotometric standard stars.

    \begin{figure*}
    \centering
    \includegraphics[width=19cm]{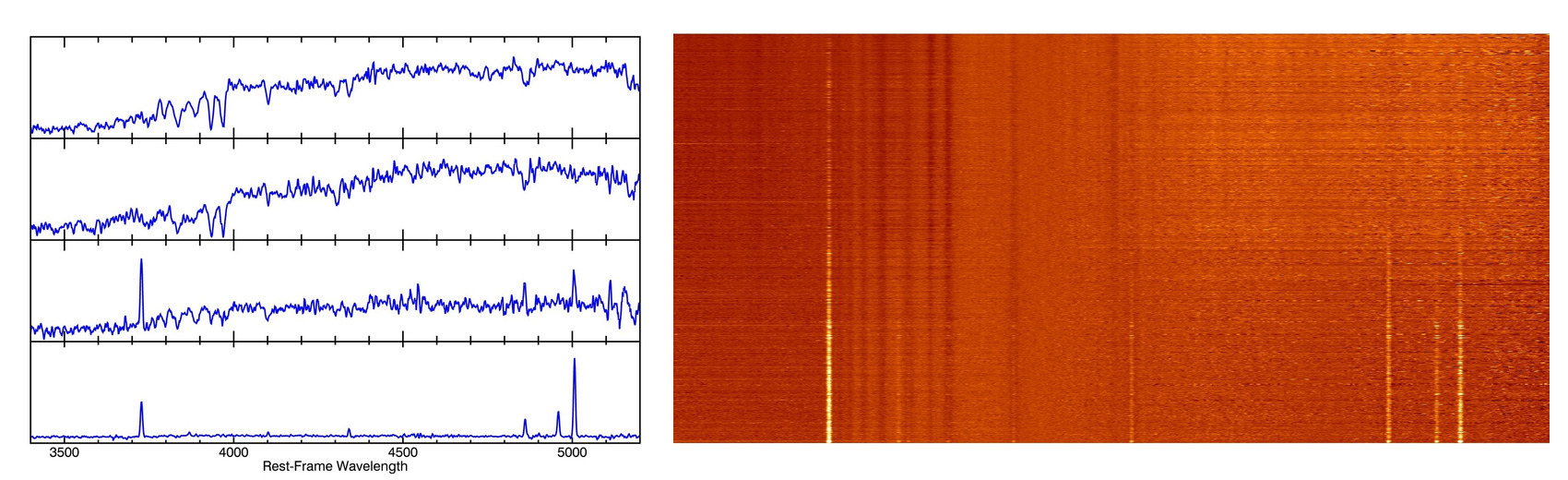}
    \caption{Left panel: examples of our typical spectra, plotted using
      rest-frame and rebinned to $2\AA$ pixel$^{-1}$. Right panel:
      montage of all our rest-frame spectra using the same wavelength range as
      for the left panel, with galaxy stellar mass decreasing from top to
      bottom.}
       \label{Fig6}
     \end{figure*}

Because of the nonstandard observing procedure used for these
observations, in which objects are observed repeatedly on different
masks, a further data reduction step has to be added to the standard
VIPGI procedure. For objects with multiple observations, a final
combination of all previously obtained 2D spectra is carried out,
accounting for the location of the target spectrum. After this final
2D combination, a 1D spectrum is extracted from the combined 2D
spectrum and is flux calibrated, following the same VIPGI procedures
outlined above.

The redshift measurement is also a multistep process. We start by
conducting a fully automated redshift determination based on
cross-correlation with template spectra coupled to emission lines and
continuum fitting, carried out with the EZ redshift measurement
software \citep{Garilli2010}. Galaxy template spectra are drawn from
those built for VVDS \citep{LeFevre2005} and zCOSMOS \citep{Lilly2007}
surveys. This automated step is followed by a visual examination of 1D
and 2D spectra of every object to assess the validity of the automated
redshift measurement, or, if needed, to compute a new redshift, based
on the wavelengths of the recognized spectral features.  A visual
examination is carried out by two people independently, while a third
person has the responsibility of establishing the final redshift
measurement associated with each spectrum together with a redshift
confidence class. These confidence classes are similar to those
previously adopted for the VVDS, zCOSMOS, and VIPERS surveys
\citep[see][]{LeFevre2005, Lilly2007, Guzzo2014}, and go from 0 when
no reliable spectroscopic redshift measurement is possible, to 4 for a
high confidence, highly secure redshift, based on a high $S/N$
spectrum and supported by obvious and consistent spectral features.
The statistical meaning of these quality flags is quantified in an
objective way in Section~\ref{errorredshift}, taking advantage of the
objects for which redshift is measured twice.

We also add to these integer confidence flags a decimal digit, reflecting an
agreement between the spectroscopic and photometric redshift measurement.  For
objects with $|z_{spec} - z_{phot}|/(1+z_{spec}) \leq 0.08, $ we add $0.5$ to
the value of the confidence flag, and otherwise we add $0.1$.  The assumed
value of $0.08$ corresponds to a conservative estimate for error in
photometric redshifts (see Section~\ref{errorphotz}). Thus, a value of $*.5$
in this decimal flag improves our confidence in the spectroscopic redshift
estimate for a few targets with poor quality spectra.

Table~\ref{tab:Tab2} lists, for primary targets of our observations, the
numbers of observed objects, the number of successfully measured redshifts
(\ie $z_{flag} \ge 1.5$), and the final number of galaxies observed within the
Wall volume, consisting of those galaxies with $z_{spec}$ in the $[0.69:0.79]$
range. The choice of including galaxies with $z_{flag} = 1.5$ in our analysis
is justified by the high probability that these redshifts are correct for
this redshift confidence class, as estimated in Section~\ref{errorredshift}.

In the table, we also list a number of secondary targets with reliable
redshifts within the Wall volume. As expected, out of a total of 120
secondary targets, for which we were able to measure a reliable
redshift, only 20 are inside the Wall volume. The same statistic is
also split into three different magnitude ranges. The number in
parentheses indicates, for each sample, the subset of targets from the
$WPS_{spec}$ sample, which are used as fillers in our slit positioning
strategy or, in the case of secondary targets, just enter the slit of
a primary target by chance.

Combining our new measurements with the \20K, we obtain a total sample
of 1277 galaxies located within the Wall volume, thus doubling the
number of galaxies within the same volume originally available from
the \20K.  Out of 1277, 619 are observed during our campaign using the
$WPS_{phot}$ sample, 237 are observed during our campaign using the
$WPS_{spec}$ sample, and 421 are taken from the \20K.  Hereinafter, we
refer to this sample as the COSMOS Wall sample.

On the left panel of Figure ~\ref{Fig6} we show some typical
spectra. The spectra are all converted to the rest-frame wavelengths,
and rebinned to $2\AA$ pixel$^{-1}$. On the right panel we show all
our rest-frame spectra, sorted in decreasing galaxy stellar mass from
top to bottom. The progressive blueing of the galaxy population moving
from higher to lower masses is readily visible, as manifested by the
progressive appearance of emission lines.

\subsection{Errors in redshift measurements}
\label{errorredshift}

We estimate a typical accuracy in our redshift measurements taking
advantage of the set of 237 galaxies from the \20K that were
reobserved during our campaign.

\begin{figure}
    \centering
    \includegraphics[width=7.1cm, angle=270]{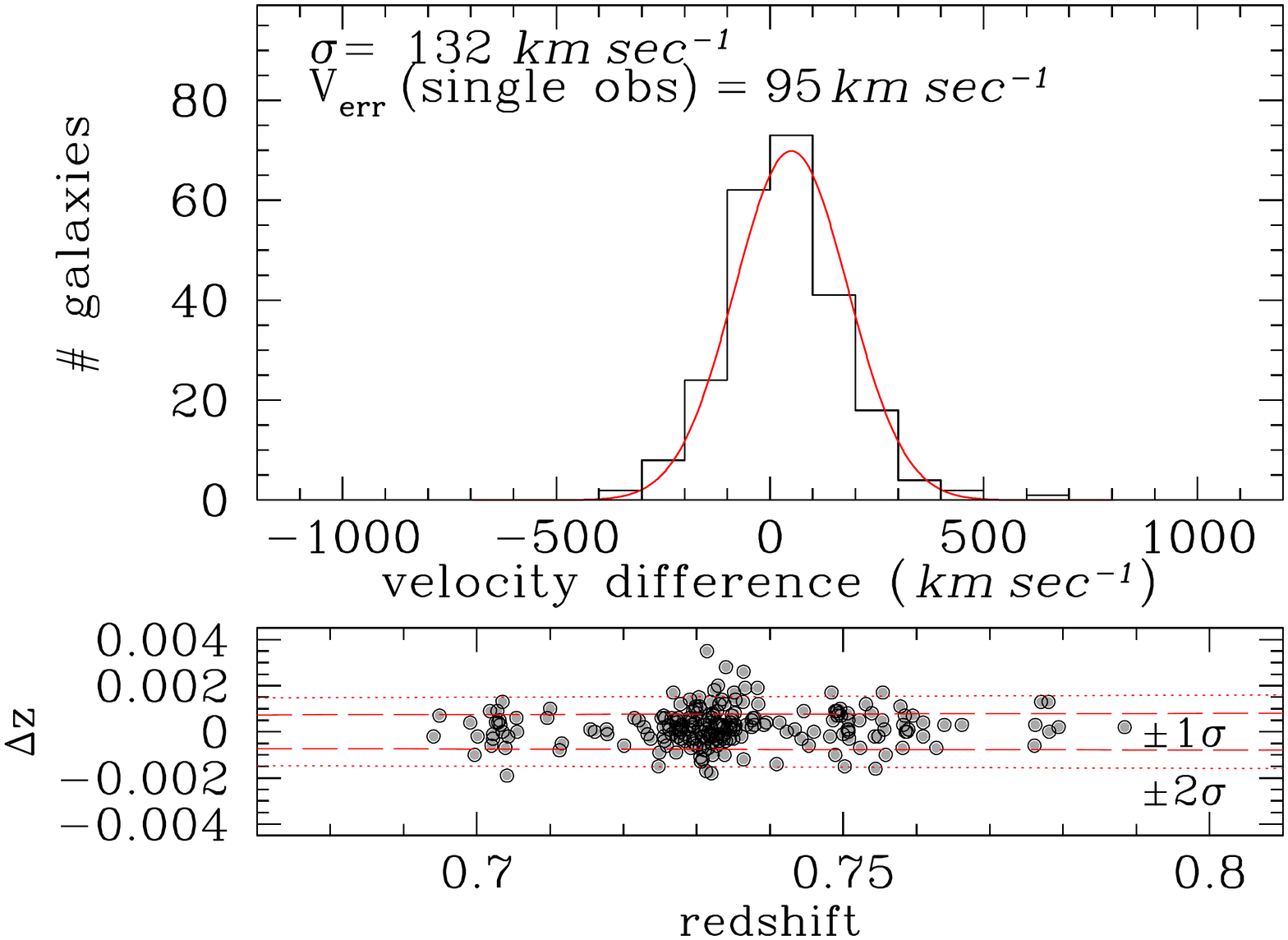}
    \caption{This plot shows our accuracy in the redshift measurements. The top
      panel shows the histogram of the distribution of velocity differences
      $\Delta v = c\times\Delta z/(1+z)$ for the set of 237 galaxies from the
      $WPS_{spec}$ that were observed twice and whose redshift was
      independently measured. Such histogram is well fitted by a Gaussian of
      width $\sigma = 132~$km\,sec$^{-1}$. The lower panel shows the
      redshift differences; the dashed (dotted) line corresponds to $\pm 1 (2)
      \sigma$ redshift difference.}
       \label{Fig7}
     \end{figure}

The top panel of Figure~\ref{Fig7} shows the distribution of the
velocity differences between  the original \20K measurements,
$\Delta v = c\times\Delta z/(1+z)$, and ours.  Such differences are well
represented by a Gaussian of width $\sigma = 132$\,km\,sec$^{-1}$.
The lower panel of Figure~\ref{Fig7} shows the redshift differences
and the dashed (dotted) line corresponds to $\pm$ 1 (2)$~\sigma$
velocity differences.  Assuming that the budget in velocity errors is
equally distributed between the \20K and our new data, we infer that
the $rms$ of the velocity accuracy of our measurements is $\sim
95$\,km\,sec$^{-1}$. In fact, the global velocity uncertainty of \20K
is estimated to be slightly higher, $\sim110$\,km\,sec$^{-1}$; see
\citet{Lilly2009}. Adopting this value for the contribution of the
\20K to the redshift differences of Figure~\ref{Fig6} results in a
lower value for the error budget of our velocity measurements of $ \sim
75$\,km\,sec$^{-1}$.  We adopted a more conservative value of $ \sim
90~$km\,sec$^{-1}$, or $\sigma_{z} = 0.0003\times(1+z)$, as an
estimate of our error in velocity measurements.

The sample of reobserved galaxies can be further used to quantify the
significance of the quality flags introduced in
Section~\ref{data_reduction}, following the simple strategy adopted
in, for example, \citet{LeFevre2005} and \citet{Guzzo2014}.  We make a
reasonable assumption that when two measured redshifts are in
agreement they are both correct and that two measurements of flag $a$
and $b$, whose intrinsic probability of being correct is $p_{a}$ and
$p_{b}$, respectively, have a probability of being concordant, thus
both correct, equal to $p_{a}\times p_{b}$.

Out of the 144 pairs of measurements whose assigned flag was 3 or 4,
both according to our classification and in the \20K original spectral
classification, no pair has a difference in redshift measurement
greater than $3\sigma_{z}\sim0.002$, corresponding to a very high
confidence level for this quality class, that is, basically $\sim
100\%$.

Out of the 45 pairs of measurements where one measurement has flag
equal 2 and the other flag equals 3 or 4, only 1 displays a difference
greater than $3\sigma_{z}$, while out of the 13 pairs of measurements
whose assigned flags are 2 for both measurements, no pair displayed a
difference in redshift measurement greater than $3\sigma_{z}$. We
therefore estimate the confidence level for this quality class as
$\sim 98\%$.

We finally estimate a confidence level for the measurements whose assigned
flag is 1.5 (galaxies whose spectroscopic redshift estimate is
uncertain, but is backed up by a good agreement with the galaxy photometric
redshift).  Out of the 20 pairs of measurements, in which one measurement has flag
equal to 1.5 and the other flag is equal to 2/3/4, four displayed a difference in
redshift measurement greater than $3\sigma_{z}$, while out of the 3 pairs of
measurements whose assigned flag was 1.5 for both, 1 has a difference in
spectroscopic redshift measurement between the two estimates that is greater than
$3\sigma_{z}$. Therefore, the confidence level for this quality class can be
safely set to $\sim 80\%$, enough to grant the use of this flag in our
subsequent analysis.

\subsection{Consistency with photometric redshifts} 
\label{errorphotz}

We use the full sample of galaxies we observed  with $z_{flag} \geq
2$ to estimate the typical error of the photometric redshift as a
function of target apparent $I$-band magnitude. This estimate is
used in our weighting scheme to account properly for the biases
introduced by the photometric preselection of our targets. Using only
galaxies with $z_{flag} \geq 3$ does not significantly impact  the
results, confirming the high reliability of galaxies with $z_{flag} =
2$ (see previous Section).

\begin{figure}
    \centering
    \vspace{-1.0cm}
    \includegraphics[width=7.5cm, angle=270]{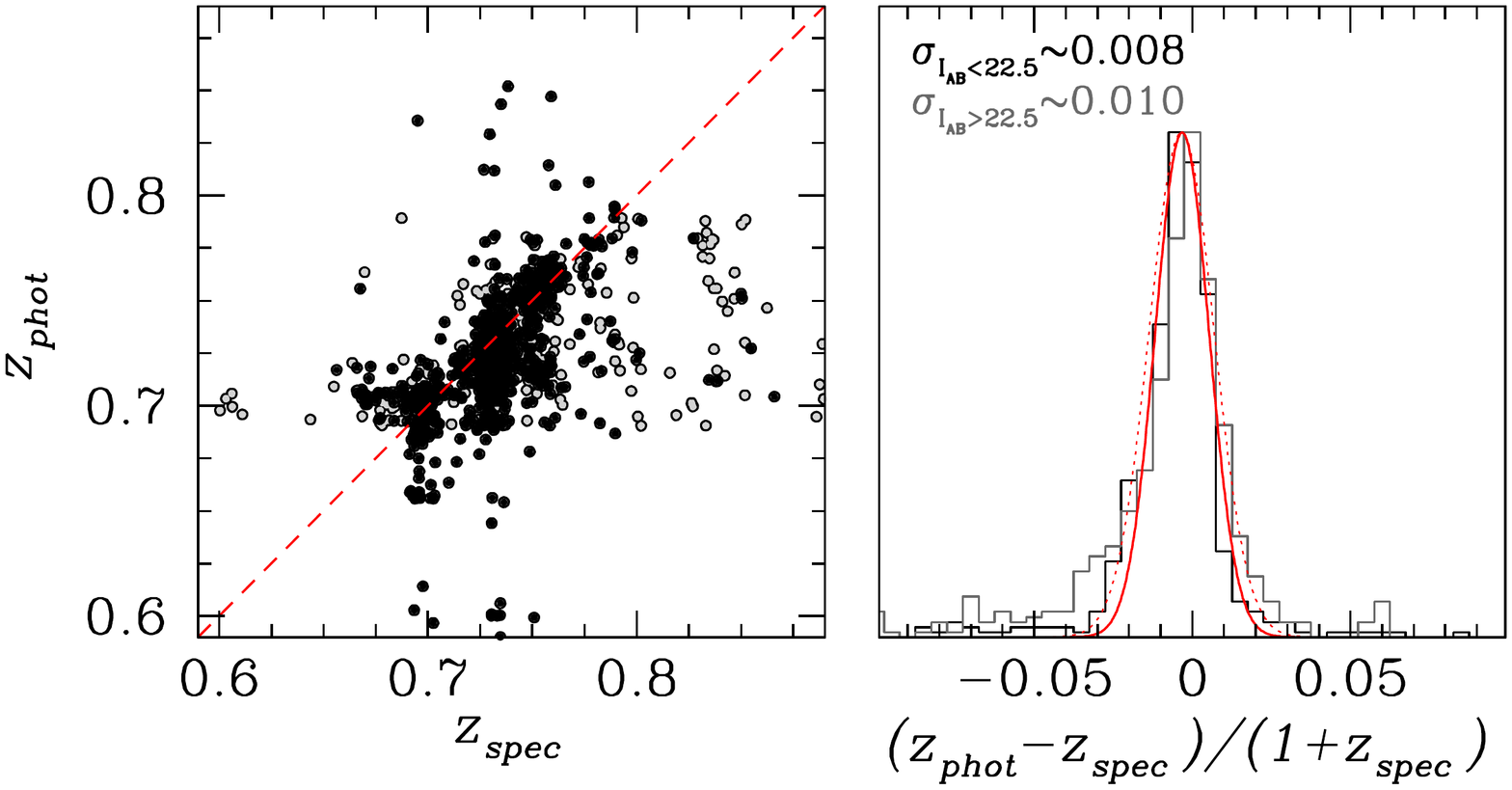}
    \vspace{-1.0cm}
    \caption{Left panel shows $z_{phot}$ vs. $z_{spec}$ for a set
      of galaxies with $z_{flag} \geq 2$, showing the
      galaxies brighter/fainter than $I_{AB}=22.5$ in black/gray. Right panel
      shows the histogram of $(z_{phot}-z_{spec})/(1+z_{spec})$ for the
      same set of galaxies and its Gaussian fit. The black/gray
      histogram refers to galaxies that are brighter/fainter than
      $I_{AB}=22.5$, while the continuous/dotted red line shows the
      Gaussian fit to the histogram of the brighter/fainter galaxies. The
      labels indicate the value of $\sigma$ for each Gaussian.}
       \label{Fig8}
     \end{figure}

For galaxies at $I_{AB} \leq 22.5$ (black points and red line in
Figure~\ref{Fig8}) the value of $(z_{phot} - z_{spec})/(1+z_{spec})$ is well
fit by a Gaussian of $\sigma \sim 0.008\times(1+z)$, which is in good agreement with
\citet{Ilbert2009}. 

For fainter sample of galaxies at $I_{AB} > 22.5$ (gray points and red
dashed line in Figure~\ref{Fig8}), the difference between photometric
and spectroscopic redshift estimates is well fit by a Gaussian with a
slightly higher value of $\sigma \sim 0.010\times(1+z)$, which is again in good
agreement with the values quoted for similarly faint samples in
\citet{Ilbert2009}.  The fraction of catastrophic errors (i.e.,
objects with $|z_{phot} - z_{spec}|/(1+z_{spec}) > 0.15 $) is always
$\sim1\%$ and agrees well with that quoted in \citet{Ilbert2009}.

Both for the brighter and fainter sample in the [0.69-0.79]
redshift range we detect a small but systematic offset between
photometric and spectroscopic redshift values on the order of
$(z_{phot}-z_{spec})/(1+z_{spec}) \sim -0.003$. This effect is visible
in the top panel of Figure~\ref{Fig2} as a slight offset between the
position of the peak of the COSMOS Wall structure in the spectroscopic
or in the photometric redshift distributions.

With our new spectroscopic redshift measurements we can also estimate
the net loss of galaxies within the Wall volume due to the use of
photometric redshift preselection, both for the brighter subset of
galaxies and for the fainter subset.

For brighter galaxies, at $I_{AB} \leq 22.5$, there are 289 galaxies
inside the Wall volume out of 333 galaxies with secure redhift from
the $WPS_{phot}$ sample, that corresponds to a global success rate of
$\sim 86\%$ and is in good agreement with the estimate obtained using
the \20K (see Section~\ref{photzselection}).

On the other hand, for fainter galaxies, at $I_{AB} > 22.5$, there are
323 galaxies inside the Wall volume out of 418 galaxies with secure
redhift from the $WPS_{phot}$ sample, that corresponds to a lower
global success rate of $\sim 77\%$, as expected.

In both cases the exact value of the global success rate is a few
percent lower than the value one would naively expect in the presence
of catastrophic photometric redshift errors as low as $\sim1\%$ and
perfectly Gaussian statistics. In fact we observe a slightly higher
percentage of discrepant objects, with respect to a Gaussian model,
whose error in photometric redshift estimate is larger than $3\sigma$
and this in turn causes a further lowering of the true global success
rate.


\section{Correcting for survey incompleteness: Weighting scheme}
\label{weightingscheme}

In this section we present a weighting scheme adopted to reconstruct a
statistically complete sample starting from our observed Wall sample
catalog.

In order to achieve this goal we need to correct for three main
factors. First, most of the targets of the parent sample are
preselected using photometric redshifts; second, not all the targets
of the parent sample are positioned on the slit; and third, we do not
succeed in measuring a redshift of every observed object.  Our
weighting scheme therefore includes three different contributions: the
photometric redshift success rate for selection, the target sampling
rate, and the spectroscopic success rate. In addition we also
introduce a correction to take into account that the target sampling
rate is spatially inhomogeneous from the widely varying surface number
density of our targets and the finite number of slits per quadrant of
each pointing in our observations. Given the negligible number of
truly new secondary targets (only seven galaxies are secondary targets
that were in the $WPS_{phot}$ sample) in the following, we do not make
any distinction in treatment between the primary and secondary
targets. In the next sections, we discuss in detail the strategy that
we adopt to correct for each of the above-mentioned factors.

\subsection{Photometric redshift success rate}
\label{PhSR}

The first correction only needs to be applied to the targets selected
from $WPS_{phot}$, which we call it photometric redshift success rate
(PhSR).

\begin{figure}
    \centering
    \includegraphics[width=7.5cm, angle=0]{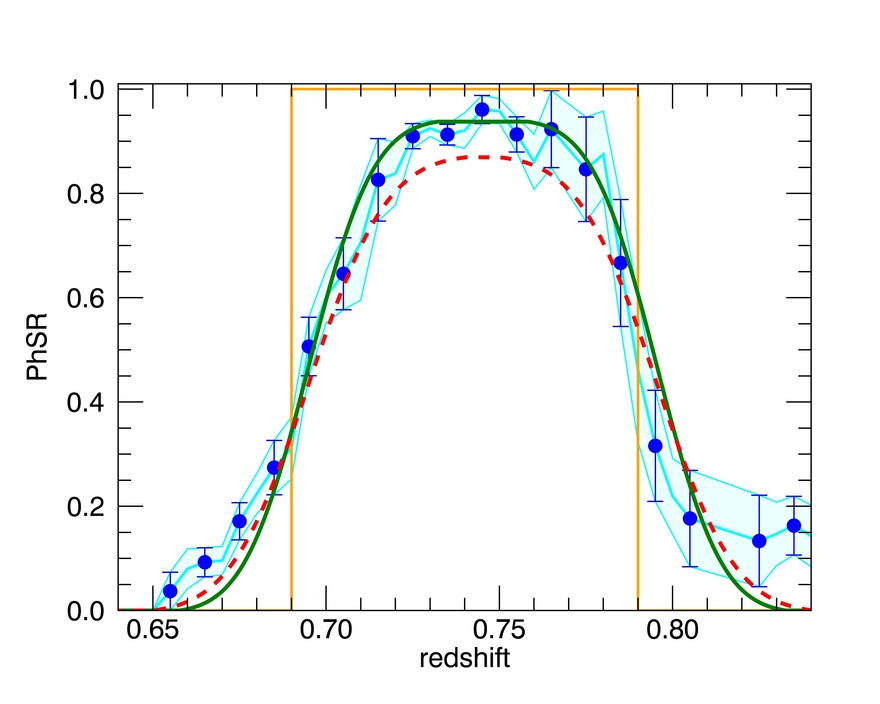}
    \caption{This plot shows PhSR as a function of redshift.  The top
      hat function plotted in orange is the selection function holding
      for $WPS_{spec}$. The shaded area and blue points refer to the
      observational estimate (using the \20K) of the success rate of
      our photometric redshift selection. The continuous green (dashed
      red) line shows analytical predictions of PhSR for the sample of
      targets within $WPS_{spec}$ that are brighter (fainter) than
      $I_{AB} = 22.5$. }
       \label{Fig9}
     \end{figure}

Targets from $WPS_{spec}$ were selected using their spectroscopic
redshift within the \20K, and thus their selection function is well
represented by a top hat function with height equal to $1$ within the
limits $0.69 \leq z \leq 0.79$ and equal to $0$ elsewhere, as
indicated by the continuous orange line in Figure~\ref{Fig9}.  

On the contrary, targets from $WPS_{phot}$ were selected using their
photometric redshift information. In the bottom panel of
Figure~\ref{Fig2} we plot the spectroscopic redshift distribution
(orange shaded histogram) for the \20K galaxies possessing a
photometric redshift in the range $0.69\leq z_{phot}\leq 0.79$, \ie
within the limits of the Wall volume according to their photometric
redshift. Clearly we deplete the true spectroscopic redshift
distribution within the Wall volume by a factor that increases
progressively when moving from the middle of our photometric redshifts
selection window toward its edges.

To better visualize this effect in Figure~\ref{Fig9} the blue points
and the cyan shaded region show the ratio, as a function of redshift,
between the shaded orange and the gray/red histograms shown in the
bottom panel of Figure~\ref{Fig2}. We consider this an estimate of the
success rate of our photometric redshift selection in recovering the
actual spectroscopic redshift distribution provided by the \20K. The
blue points are in bins of $\Delta z = 0.01$, while the error bars are
obtained using the formula for binomial errors as in
\citet{Gehrels1986}; the cyan region is obtained with a running bin of
width $\Delta z = 0.01$ in steps of $\delta z = 0.005$ and similarly
binomial error estimates.

The success rate of the photometric redshift selection (cyan region)
has a strong dependence on redshift: it is more than $90\%$ for galaxies
at the center of our photometric selection window while it drops to lower
values near the edges of the selection window.

Such a dependence can be estimated analytically using the results on
photometric redshift errors, as discussed in Section~\ref{errorphotz}.
For targets brighter (fainter) than $I_{AB} = 22.5$ the typical
photometric redshift error includes a systematic offset
$\delta_{offset}=(z_{phot}-z_{spec})/z_{spec} \sim -0.003~(-0.003)$
and a dispersion $\sigma_{photz} \sim 0.008~(0.010) \times(1+z)$,
together with a global success rate of $\sim 86~(77)\%$ within the
Wall volume. We can then analytically parametrize  the observed PhSR as
a function of redshift by making the convolution of the ideal top hat
selection function with a Gaussian of dispersion equal to
$\sigma_{photz}$, further adding a systematic offset equal to
$\delta_{offset}$, and finally renormalizing the integral of the
resulting function within the redshift range [0.69-0.79] to the actual
value of the global success rate estimated within the Wall volume.

The agreement between the observed and the parametrized PhSR(z) is excellent, as shown
by the cyan region and green continuous line in
Figure~\ref{Fig9}, respectively.  We can therefore confidently use
this procedure to compute  the PhSR analytically both for the brighter
and fainter subsets of our targets using the appropriate values
for photometric redshift errors as derived in
Section~\ref{errorphotz}. The red dashed line in Figure ~\ref{Fig9}
shows the results for the fainter subset of our targets.

We adopt these two function as our parametrization of PhSR(z) of
brighter/fainter targets from $WPS_{phot}$, while for targets from
$WPS_{spec}$ the PhSR is simply equal to $1$.  The weight associated with the
PhSR is: $w_{PhSR}(z) = 1/$PhSR(z).

   \begin{figure}
    \centering
    \vspace{-0.8cm}
    \includegraphics[width=6.5cm, angle=270]{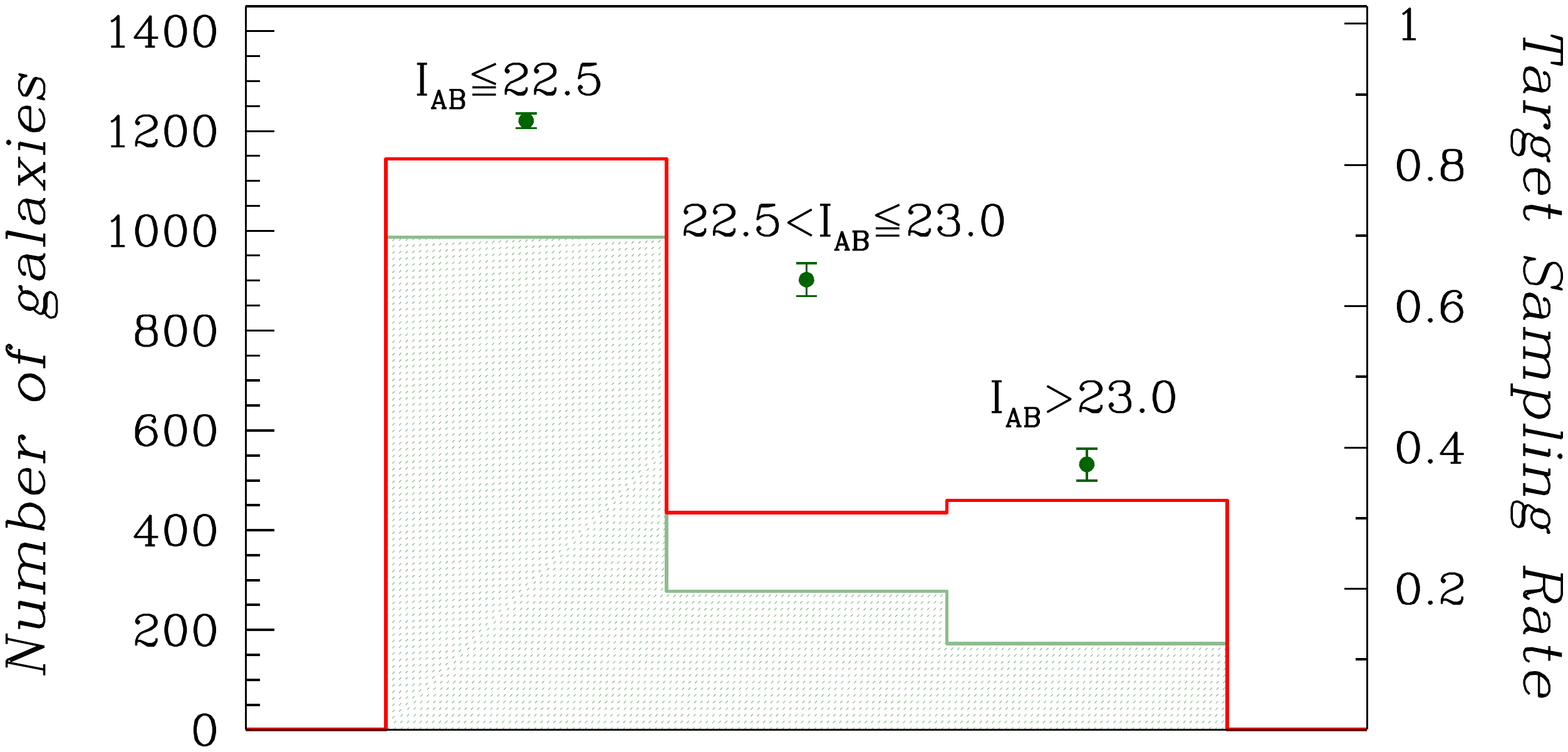}
    \caption{TSR in the three $I_{AB}$ magnitude bins, used to define
      our slit positioning strategy: $I_{AB}\leq 22.5$,
      $22.5<I_{AB}\leq 23.0$, $I_{AB}>23.0$. The magnitude bin
      considered is labeled atop of each bin. The red histogram shows
      the magnitude distribution of the targets in the total Wall
      parent sample. The filled green histogram shows the magnitude
      distribution of the targets where we could position on a slit
      (for both histograms the correponding number of targets is shown
      on the left y-axis). The filled points are the TSR, which
      is ratio between the two histograms, with values given by 
      the right y-axis.}
       \label{Fig10}
     \end{figure}

\subsection{Target sampling rate}
\label{TSR}

The target sampling rate (TSR) accounts for the selection function of
the photometric sources targeted by observations, as we did not
position a slit on all the galaxies in the parent sample. The TSR is
thus defined as the ratio between the number of photometric sources in
the parent sample and the number of photometric sources we targeted
with VIMOS observations.

By parent sample, unless specified otherwise, we always mean the total
parent sample, that is the sum of $WPS_{phot}$ and
$WPS_{spec}$. Obviously all the objects of $WPS_{spec}$ already
possess a spectroscopic redshift from \20K data so all its galaxies
have been {\it observed} by definition. In the computation of the TSR
(and later of the spectroscopic success rate), these objects are
treated together with those of the $WPS_{phot}$ to constitute the
total parent sample; we consider these among the set of targets where
a slit was positioned and a successful redshift was measured. The only
difference for the objects of $WPS_{spec}$ is in the treatment of
their PhSR, as discussed in the previous Section.

As our observing strategy is based on the target $I_{AB}$ magnitude
(see Section~\ref{ObsStrategy}), we parametrize the TSR as a function
of targets $I_{AB}$ magnitude in the three $I_{AB}$ magnitude bins
used to define our slit positioning strategy, that is $I_{AB}\leq 22.5$,
$22.5<I_{AB}\leq 23.0$, $I_{AB}>23.0$.

In Figure~\ref{Fig10} we plot in red the distribution of the
number of targets in the parent sample within these three magnitude
bins, while the filled green histogram refers to the magnitude
distribution of the targets positioned on a slit. The filled points
show the ratio between the two histograms, that is the estimate TSR in
our three bins of $I_{AB}$ magnitude. The TSR errors are computed
using the formula for binomial errors as in \citet{Gehrels1986}.  Our
average TSR is $ \sim 70\%$, taking into account the \20K data.
The weight associated with the TSR is $w_{TSR}(I_{AB}) = 1/$TSR$(I_{AB})$.

   \begin{figure}
    \centering
    \vspace{-3.2cm}
    \includegraphics[width=10.2cm, angle=0]{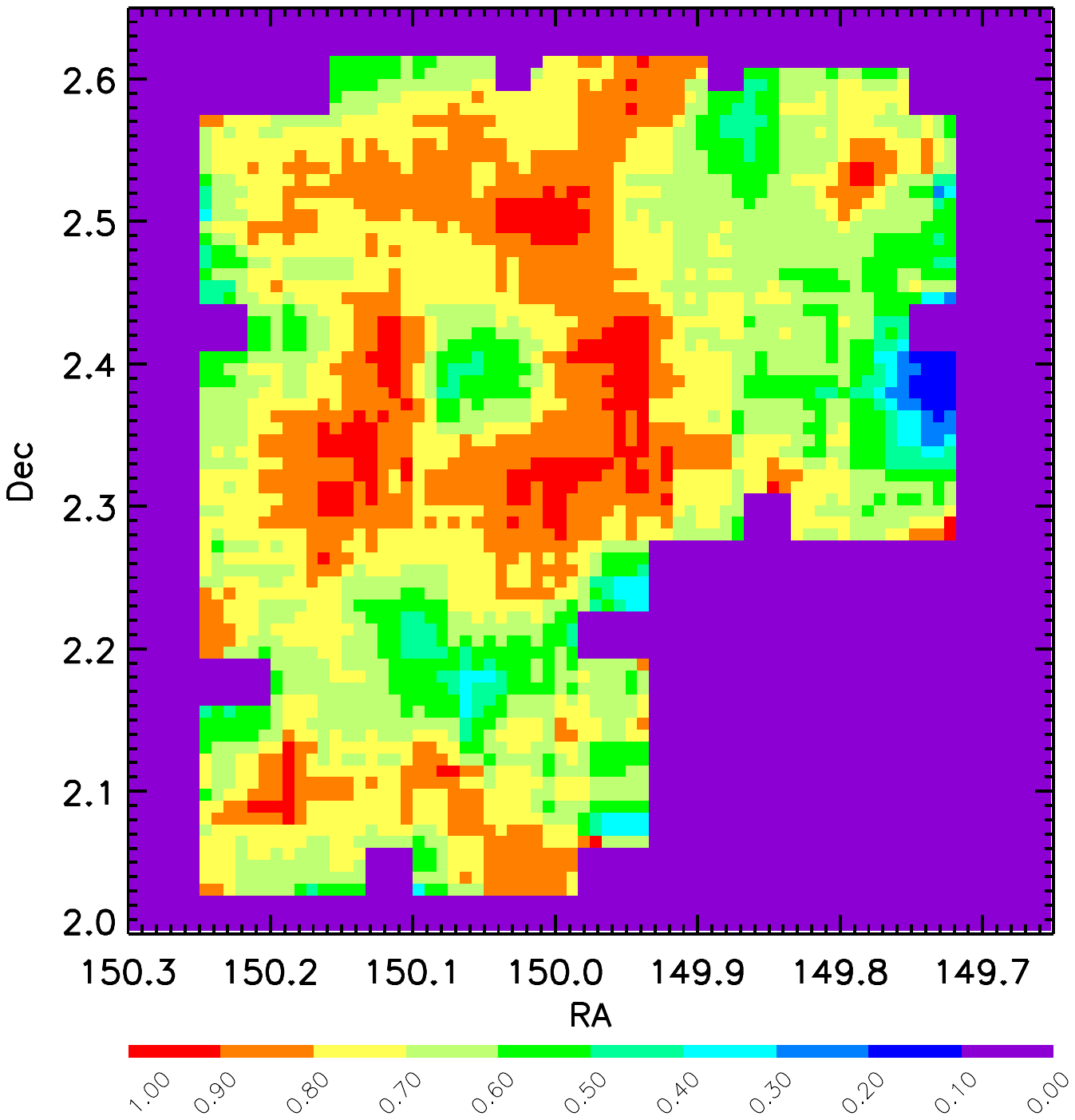}
    \vspace{-3.1cm}
    \caption{RaDec Sampling Rate before normalization, showing the
      spatial variation of the TSR due to the sky inhomogeneities in
      our survey coverage. Comparing this Figure with
      Figure~\ref{Fig5} it is possible to see that the highest density
      region where the richest structure is located suffers from a
      slight relative under-sampling, despite having been targeted by
      four different pointings.}
       \label{Fig11}
     \end{figure}

\subsection{RA-Dec inhomogeneities of TSR}
\label{radecSR}

There is a further spatial modulation on the TSR that needs to be
taken into account when estimating those physical quantities that
depend on the inhomogeneities in our survey coverage.  Typical
examples of these quantities, such as local densities or group
richness estimates.

In this case we need to introduce a RA-Dec sampling rate (RaDecSR), to
take into account that the target sampling rate is spatially
inhomogeneous owing to widely varying surface number density of our
targets and the finite number of slits per quadrant of each pointing
in our observations.

To compute RaDecSR we adopt a strategy similar to that one of
\citet{Iovino2010}, except that we adopt a slightly wider smoothing
box size given the lower surface density of our parent catalog. In a
grid of steps equal to $30\arcsec$ in right ascension and declination,
and in square cells of $4\arcmin\times4\arcmin$, we compute the ratio
of the number of targets where a slit is positioned to the total
number of parent sample targets within the same area.  We then obtain
RaDecSR($\alpha,\delta$) by normalizing to unity the mean value of
this ratio over the full $RA-Dec$ coverage of the Wall. The weight
associated with RaDecSR is $w_{RaDecSR}(\alpha,\delta) =
1/$RaDecSR($\alpha,\delta$).

In Figure~\ref{Fig11} we show the spatial variation of
RaDecSR($\alpha,\delta$) before normalization; the color scale
provides the legend for the range of values shown on the plot. One
can appreciate that despite having been targeted with four different
pointings, the area where the largest cluster is located still has a
relative under-sampling due to the high local density of targets.

\subsection{Spectroscopic success rate}
\label{SSR}

Finally, the spectroscopic success rate (SSR) accounts for the fact
that not all the sources where we positioned a slit provide a
successful redshift measurement. The SSR is thus defined as the ratio
between the number of photometric sources where we positioned a slit
and the number of reliable redshift measurements obtained. Given the
good quality of most of our spectral data, the total SSR is on the
order of $\sim 95\%$, and we adopt a simple parametrization based on
the target $I_{AB}$ magnitude, as shown in Figure~\ref{Fig11}. A more
complex functional form for the SSR, which also includes the
rest-frame galaxy colors, is deemed unnecessary (see also Figure~2 in
\citet{Lilly2009}, where no significant dependence on the rest-frame
galaxy colors for the fraction of spectra yielding a successful
redshift measurement is visible at $z\sim 0.7$).

   \begin{figure}
    \centering
    \vspace{-0.8cm}
    \includegraphics[width=6.5cm, angle=270]{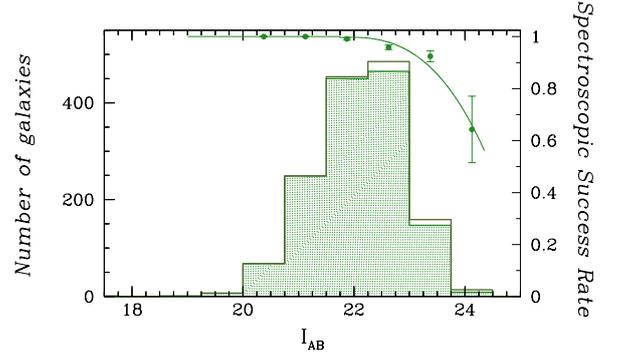}
    \caption{SSR as a function of a target $I_{AB}$ magnitude.  The
      dark green histogram shows the magnitude distribution of the
      targets from the total parent sample where a slit was
      positioned. The filled light green histogram shows the
      magnitude distribution of the targets whose redshift is reliably
      measured (for both histograms the numbers are indicated on the left
      Y-axis). The filled points are the SSR, which is a ratio of the
      two histograms whose value is indicated on the right
      Y-axis. The continuous light green curve is the adopted function
      to parametrize SSR($I_{AB}$).}
       \label{Fig12}
     \end{figure}

In Figure~\ref{Fig12} the dark-green histogram shows the magnitude
distribution of the targets from the total parent sample where a slit
was positioned, while the filled light green histogram shows the
magnitude distribution of the targets whose redshift was reliably
measured (for both histograms numbers are indicated on the left y-axis
values). The filled points indicate the SSR, computed as a ratio of
the two histograms. The continuous light green curve is the function
adopted to parametrize SSR($I_{AB}$).  The weight associated with SSR
is $w_{SSR}(I_{AB}) = 1/SSR(I_{AB})$.

For each galaxy $i$ we can thus estimate its total weight using the formula
$w_i = w(\alpha_i, \delta_i, z_i, I_{ABi})$ = $w_{PhSR}(z_i)\times
w_{TSR}(I_{AB\;i}) \times w_{RaDecSR}(\alpha_i,\delta_i) \times
w_{SSR}(I_{AB\;i})$.


\section{Rest-frame magnitudes and stellar masses} 
\label{Physquantities}

The spectroscopic data of the new COSMOS Wall sample of galaxies is
matched with the multiwavelength photometric data for each galaxy to
derive rest-frame galaxy properties. We then perform SED fitting to
obtain luminosities and stellar masses for each galaxy in our sample
using an updated version of {\it Hyperzmass}
\citep{Bolzonella2000,Bolzonella2010} and a strategy similar to that
adopted in \citet{Davidzon2013}.

Given a set of synthetic SEDs, {\it
  Hyperzmass} fits these models to the multiband photometry for each
galaxy and subsequently selects the model that minimizes the $\chi^2$.
We compiled a set of synthetic SEDs using simple stellar populations
(SSPs) provided by the models of \citet{BruzualCharlot2003}, adopting
the \citet{Chabrier2003} IMF, and assuming a nonevolving stellar
metallicity, for which we consider two values: solar ($Z=Z_{\sun}$)
and subsolar ($Z=0.2~Z_{\sun}$).  We assume an exponentially
declining star formation history (SFH), for which SFR $\propto exp(-
t/\tau)$, with the timescale $\tau$ ranging from 0.1 to $30$ Gyrs.

Finally for the galaxy dust content, we implement the
\citet{Calzetti2000} and Prevot-Bouchet \citep{Prevot1984,Bouchet1985}
extinction models, with values of $A(V)$ ranging from 0 (no dust) to 3
magnitudes.

\citet{Davidzon2013} provide for a lengthy discussion of possible
alternative choices of codes and spectral library parametrizations and
their impact on stellar mass estimates.  Here, it is enough to mention
that given the wide range of physical properties allowed in the SED
fitting procedure, we decided to exclude some unphysical parameter
combinations from the fitting.
In particular, we limit the amount of dust in passive
galaxies (i.e., we impose $A(V) \leq 0.6$ for galaxies with $age/\tau
> 4$), we avoid very young extremely star-forming galaxies with short
$\tau$ timescales (i.e., we prevent fits with models with $\tau \leq
0.6$ Gyr when requiring $z_{form} < 1$), and we only allow ages to be
between 0.1 Gyr and the age of the Universe at the spectroscopic
redshift of the fitted galaxy
\citep[see][]{Pozzetti2007,Bolzonella2010}.

We also derive absolute rest-frame magnitudes for each galaxy of the
COSMOS Wall sample, using the apparent magnitude that most closely
resembles the observed photometric passband, shifted to the redshift
of the galaxy under consideration. The $k-$correction factor in this
method is much less sensitive to the adopted SED template type than
using a global filter transformation that is confined to a single
specific filter passband \citep[see Appendix A of][]{Ilbert2005}.  

\subsection{Mass completeness of COSMOS Wall sample}
\label{masscompleteness}

\begin{figure}
    \centering
    \vspace{-2.0cm}
    \includegraphics[width=9.2cm, angle=0]{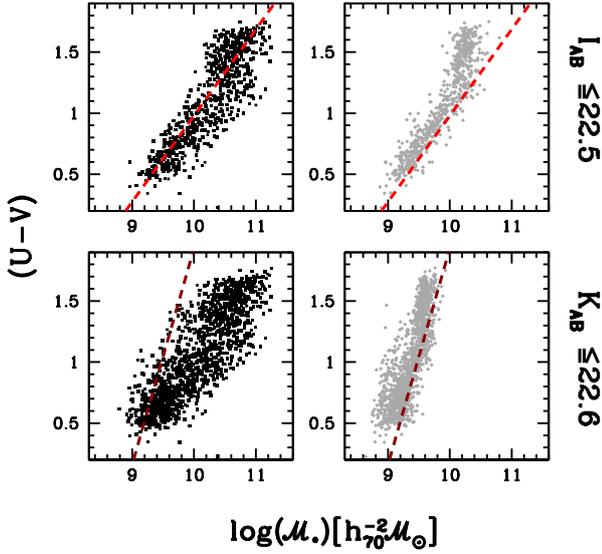}
    \vspace{-2.1cm}
    \caption{Stellar mass completeness as a function of galaxy $(U-V)$
      rest-frame colors for the whole COSMOS Wall sample limited to
      $K_{AB} \leq 22.6$ (bottom panel) and for the subset limited at
      $I_{AB} \leq 22.5$ (top panel), which reproduces the
      completeness of the \20K. Left hand panels display the actual
      galaxy stellar mass distribution, while the right hand panels
      show the limiting masses for each galaxy, all as a function of
      $(U-V)$ rest-frame colors, plotted on the y-axis.  The dashed
      lines represent the mass limits estimates. }
       \label{Fig13}
     \end{figure}

As a first test we estimate the stellar mass limits for galaxies of the COSMOS
Wall sample. We follow the prescriptions of \citet{Pozzetti2010} to compute
for each galaxy its limiting stellar mass, \ie the stellar mass it would have,
at its spectroscopic redshift, if its apparent magnitude is equal to the
limiting magnitude of our survey ($K_{AB} = 22.6$). We then use these
estimated limiting masses to define, in bins of $(U-V)$ rest-frame colors, the
mass $M^*_{cut-off}$, below which $85\%$ of limiting masses of galaxies of
that color lie. In Figure~\ref{Fig13} we plot these mass limits, together with
those of the subset having $I_{AB} \leq 22.5$, to compare our actual
gain in mass completeness with respect to the \20K  directly.

In this Figure the right panels show the values of these limiting
masses while the left panels represent the galaxy stellar mass
distribution, both as a function of rest-frame $(U-V)$ colors. The top
panels refer to the subset limited to $I_{AB} \leq 22.5$, while the
bottom panels correspond to the whole COSMOS Wall sample. The dashed
lines represent the mass limits computed as discussed above; the
increment in stellar mass depth achieved with respect to the original
\20K is readily appreciable. We confirm that we reached the initial
mass limit goal of log$({\cal M}_{*}/{\cal M}_{\odot})= 9.8$ for the
reddest galaxies of our sample, thus improving by nearly a decade in
mass the original \20K completeness limit at these redshifts
\citep{Pozzetti2010, Bolzonella2010, Iovino2010}.
 
The adopted $K_{AB }$ selection makes the stellar mass limit much less
dependent on galaxy colors than an $I_{AB}$ selection; the line that
indicates the mass limit in Figure~\ref{Fig13} is significantly less
slanted for the $K_{AB }$ selection than for the $I_{AB}$
selection. As a consequence, we can also incorporate in our analysis a
significant tail of the $I_{AB }$ bright and blue galaxy population,
which is absent in the mass complete subset of the \20K.  This is
another way to state that using $K_{AB }$ selection enables us to
observe the missing red counterparts of the existing low-mass blue
galaxies in the \20K.  Thus, with our survey we double the sample of
spectroscopic redshifts available within the Wall volume, but we
roughly quadruple the number of galaxies that are used to define a
stellar mass complete sample down to our mass limit of log$({\cal
  M}_{*}/{\cal M}_{\odot}) \sim 9.8$. In the \20K only $\sim 200$ out
of 658 galaxies in the z-range $[0.69-0.79]$ form a complete sample
down to log$({\cal M}_{*}/{\cal M}_{\odot})\geq 10.7$, while in the
new COSMOS Wall sample $\sim 800$ galaxies form a complete sample down
to log$({\cal M}_{*}/{\cal M}_{\odot})\geq 9.8$ out of a total of 1277
galaxies.

   \begin{figure}
    \centering
    \vspace{-0.2cm}
    \includegraphics[width=8.5cm, angle=0]{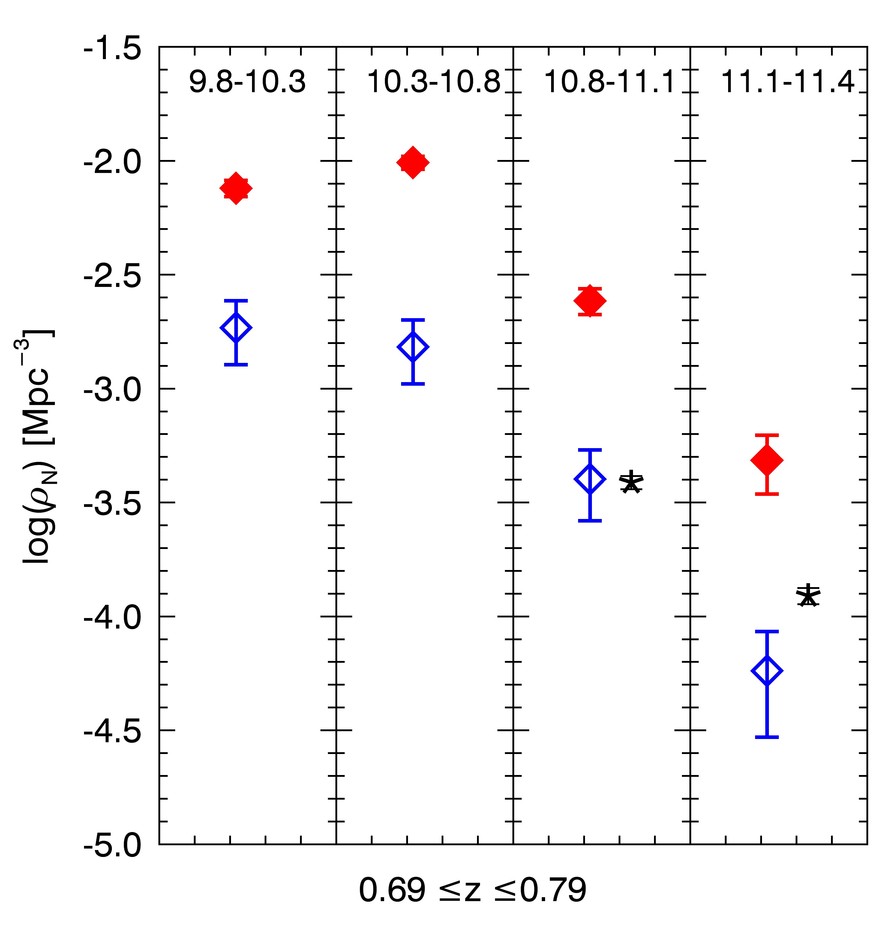}
    \vspace{-0.1cm}
    \caption{Number density of galaxies inside and outside the Wall
      structure indicate red filled diamonds and blue empty diamonds,
      respectively. The stellar mass bin considered is shown on the
      top of each panel.  The black starred points show the number
      density of galaxies within the same stellar mass bins and in the
      whole redshift range $0.69 \leq z \leq 0.79$ for the much larger
      VIPERS survey. } 
       \label{Fig14}
     \end{figure}

A simple sanity check for both our weighting scheme and our stellar mass
estimates is obtained by computing the weighted number density of galaxies
in bins of galaxy stellar mass, using for each galaxy, the weights discussed in
Section~\ref{weightingscheme}. Our results can then be compared with those
from the survey VIPERS \citep{Guzzo2014}, whose wide area coverage enables a
very precise determination of this quantity \citep{Davidzon2013}.

The outcome of this test is shown in Figure~\ref{Fig14}. We split the whole
COSMOS Wall sample in four galaxy stellar mass bins, as indicated on the top
of each panel, and into two redshift ranges: the first corresponding to the
Wall structure at $0. 72 \leq z \leq 0.74$ (filled red diamonds) and the
second to the surrounding foreground and background regions at $0.69 \leq z <
0.72$ and $0.74 < z \leq 0.79$ (empty blue diamonds). The black stars in
Figure~\ref{Fig14} are computed using the whole DR1 VIPERS survey data set in
the redshift bin $0.69 \leq z \leq 0.79$.  Error bars include both Poissonian
and cosmic variance contribution except for
the points of the Wall structure, where the only error bar shown is the
Poissonian one, as we do not claim to be dealing with a representative volume
of the universe;  see \citet{DriverRobotham2010}.

While the region of the Wall structure displays a significant
overdensity, as expected, with respect to the VIPERS data points, when
this region is excised from the analysis, the number densities agree
well with those obtained by the VIPERS data set.

In the paper \citet{Petropoulou2016a}, we will present a detailed
analysis of the galaxy stellar mass function in the Wall volume, both
as a function of environment and of galaxy colors and star formation
properties; this paper will provide a more thorough investigation of
this topic.

\section{Detection of groups} 
\label{groupsdetection}

We targeted the Wall volume because of its interesting large-scale
structure: a significant galaxy concentration and two long filamentary
structures, embedding groups, extending from it. In the original \20K
group catalog there are 19 (6) groups with 5 (10) or more
spectroscopic members within the Wall Volume, and 11 (6) are located
within the Wall structure at $0.72 \leq z_{spec} \leq 0.74$.

With the new data set available we can improve the definition of
groups as a result of our better spatial sampling within this interesting
region.  We use a friend-of-friends (FOF) group detection algorithm,
in a similar fashion to \citet{Knobel2012} and \citet{Knobel2009},
adopting the same optimization strategy as defined in these
papers. The FOF algorithm has three parameters: the comoving linking
length as defined by the {\it b} parameter, the maximum perpendicular
linking length in physical coordinates, $L_{max}$, and the ratio $R$
between the comoving linking lengths along and perpendicular to the
line of sight. \citet{Eke2004} and
\citet{Knobel2012} provide a definition of these parameters and a
discussion of their optimal values.

  \begin{figure*}
    \centering
    \includegraphics[width=20cm, angle=0]{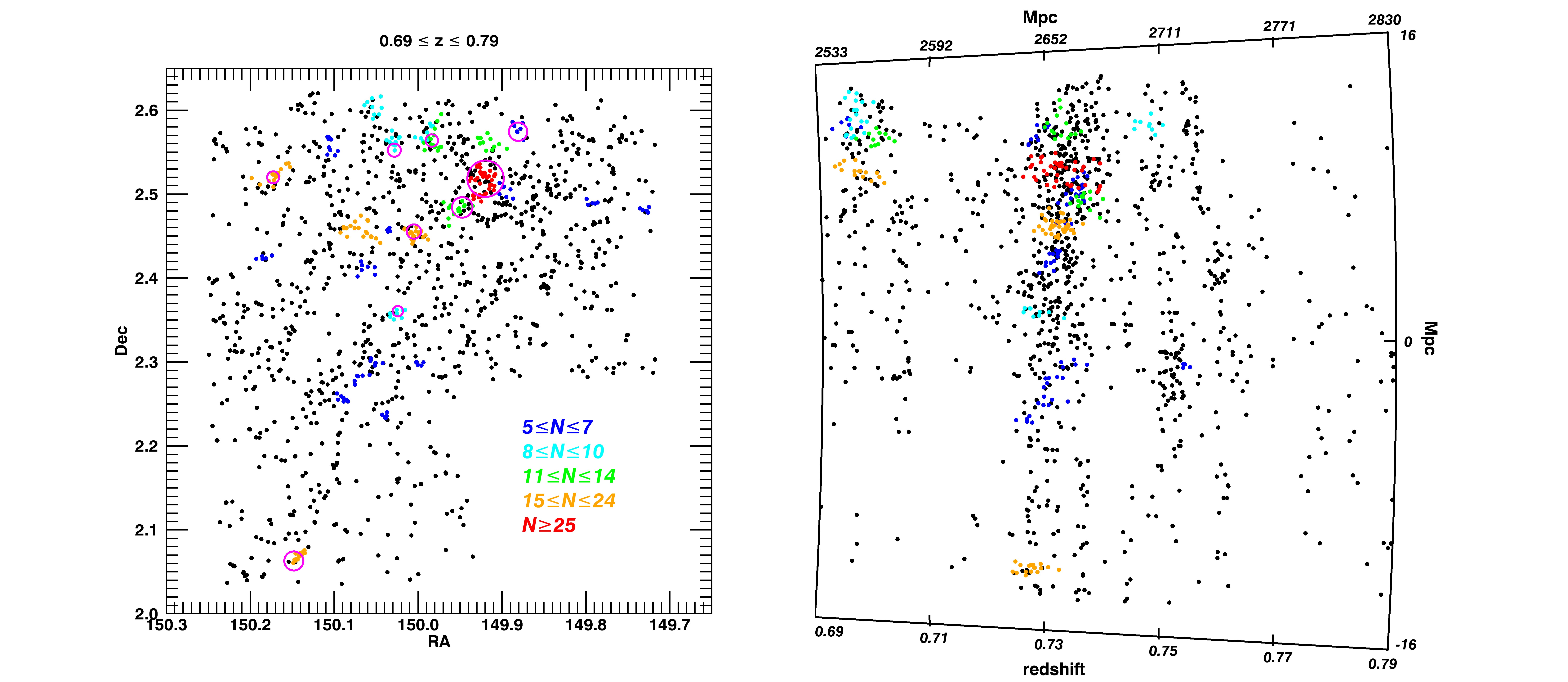}
    \hspace{-0.5cm}
    \caption{Left panel: RA-Dec distribution of the total Wall sample (black
      points) and group member galaxies, color coded as indicated in the
      legend. Large magenta open circles indicate the sky position of the
      extended X-ray sources from \citet{George2011}. Right panel: distribution of declination and redshift for the
      Wall sample and group member galaxies; color legend as in the left
      panel. Labels on the top show distances from observer in comoving
      coordinates (Mpc units), while labels on the bottom show redshift
      values. Labels on the right indicate distances in comoving coordinates
      (Mpc units) from field center along Dec direction at $z \sim 0.79$.
      For the sake of clarity, in this cone diagram we compressed
      by a factor $10$ the scale along the redshift direction.}
       \label{Fig15}
     \end{figure*}

The  comoving linking length
perpendicular to the line of sight, $l_{\bot}$, is defined as

\begin{equation}
l_{\bot} = min [L_{max}(1+z),\;b \ times \bar{n}^{-1/3}(z)]
\end{equation} 

where the free parameter $L_{max}$, expressed in physical coordinates,
prevents unphysically large values for $l_{\bot}$, when the galaxy
distribution is sampled very sparsely and $\bar{n}(z)$, the mean
density of galaxies, is low. In our survey the mean intergalactic
distance between galaxies $\bar{n}^{-1/3}(z)$ is equal to $\sim 4.8$
Mpc in the redshift range $0.715\leq z \leq 0.775$, while it
progressively increases on moving away from the central, better
sampled region of the Wall volume, reaching the values of $\sim 6.33$
Mpc at $z= 0.69$ and $\sim 5.58$ Mpc at $z=0.79$ and following our lower
sampling rate in these outer regions (see also Figure~\ref{Fig9}).

The comoving linking length parallel to the line of sight,
$l_{\parallel}$, is defined as

\begin{equation}
l_{\parallel} = R \times l_{\bot} 
,\end{equation} 
where the parameter $R$ allows $l_{\parallel}$ to be larger than
$l_{\bot}$ to take into account the elongation of groups along the line of
sight due to the so-called Fingers of God effect.

Two galaxies {\it i} and {\it j}, at comoving distances $d_i$ and
$d_j$, are assigned to the same group if their angular separation
$\theta_{\it ij}$ satisfies the condition
\begin{equation}
\theta_{ij} \leq \frac{1}{2}\left(\frac{l_{\bot,i}}{d_i}+\frac{l_{\bot,j}}{d_j}\right) 
\end{equation} 

and simultaneously, the difference between their distances satisfies
the condition 

\begin{equation}
|d_{i} -d_{j}| \leq \frac{l_{\parallel,i}+l_{\parallel,j}}{2} 
.\end{equation} 

We adopt the same multirun procedure as in \citet{Knobel2009,
  Knobel2012}, running the group detection algorithm in different
steps optimized for the different observed richness ranges, going from
the richest to the poorest groups. In each step only those groups are
accepted that have not been found at the previous steps, and the
algorithm works down in richness, as the richer groups are more easily
detected.

\begin{table}
\caption{Multirun parameter sets for FOF. $L_{max}$ is expressed in
  physical coordinates.}
\label{tab:Tab3}  
\centering                  
\begin{tabular}{c c c c c c}    
\hline\hline                 
   Step  & $N_{min}$ & $N_{max}$ & $b$   & $L_{max}$ & $R $  \\   
\hline \hline                    
   1     &  11      &  500      & 0.1   & 0.375    &  18.  \\      
   2     &   7      &   10      & 0.095 & 0.375    &  18.  \\      
   3     &   6      &    6      & 0.09  & 0.375    &  17.  \\      
   4     &   5      &    5      & 0.085 & 0.35     &  17.  \\      
   5     &   4      &    4      & 0.075 & 0.3      &  16.  \\      
   6     &   3      &    3      & 0.07  & 0.275    &  16.  \\      
   7     &   2      &    2      & 0.06  & 0.225    &  16.  \\      
\hline                             
\end{tabular}
\end{table}

Our choice of parameters follows that in \citet{Knobel2012}, as
defined after a lengthy optimization procedure, and that paper
provides more details.  The choice of the multirun parameter sets for
FOF is listed in Table~\ref{tab:Tab3}.  We detect a total of 57/26/9
groups with 3/5/10 or more observed spectroscopic members within the
Wall volume, and 34/19/6 are located within the Wall structure at
$0.72 \leq z_{spec} \leq 0.74$.

In Figure~\ref{Fig15}, we show the distribution of groups with five or
more observed members (color coded according to the number of
spectroscopic members; see legend) both in RA-Dec plane and along the
redshift direction.  Black points refer to the whole Wall
sample. Labels on top of the cone diagram represent distances in
comoving Mpc, while labels on the right show the size of the
corresponding sky region. For the sake of clarity, in the cone
shown in the right panel we stretched the
scale along the transverse direction by a factor $10$.

In Table~\ref{tab:Tab4} we list all the groups with five or more observed
members sorted according to their observed richness. For each group,
we provide its ID; the position in RA and Dec of its center, estimated
from the inverse Voronoi area weighted mean of the positions of group
members \citep[see][]{Presotto2012, Knobel2012}; its center in
redshift, estimated as the mean redshift of group members; the number
of its members possessing a spectroscopic redshift $N_{obs}$; its
corrected richness $N_{corr}$, obtained by weighting each member
galaxy with the weighting scheme discussed in
Section~\ref{weightingscheme}; and the group velocity dispersion
$\sigma_{gap}$ estimated using the gapper method \citep{Beers1990}.

As expected there are some non-negligible differences between our
group catalog and that presented in \citet{Knobel2012}; the
increased space density of our galaxy catalog makes our group search
more reliable. Some of the \citet{Knobel2012} groups are split into
smaller groups in our catalog, while the reverse is less common, in
agreement with an expected higher rate of over-merging compared to
fragmentation for the \20K group catalog, as discussed in
\citet{Knobel2012}.

The most notable case is the one of their group $\#34$ at $z\sim0.73$
with 19 spectroscopic members. This is the richest group among those
listed in \citet{Knobel2012} and is located within the Wall volume. In
our catalog this group is split into three smaller groups with four,
four, and two members respectively. It is difficult to assess if this
could indeed be a case of over fragmentation within our group
catalog. We note, however, (see also below) that no detected X-ray
emission from \citet{George2011} coincides spatially with the missing
group $\#34$ of \citet{Knobel2012} or with the smaller groups into
which we have fragmented it.

The largest group we detect, group $W1$ in Table~\ref{tab:Tab4} with
39 observed members, can also be found within the \citet{Knobel2012}
catalog, albeit in a slightly different configuration. Only 12 of the
member galaxies of group $W1$ possessed a reliable spectroscopic
redshift in the \20K, and 10 were grouped together in group $\#32$ of
\citet{Knobel2012}.

In general, if we consider a subset of the galaxies in the Wall volume
that we deem isolated (i.e., not members of any of our groups,
including pairs, triplets, and quartets) and that were observed in the
\20K, only $\sim 15\%$ of these were classified as group members in
\citet{Knobel2012}. This number further decreases by a factor of two
if we consider only those classified in \citet{Knobel2012} as members of groups with
multiplicity above 3. Vice versa, if we consider
the subset galaxies in the Wall volume classified as members in our
whole group catalog, such that at least two group members were
observed in the \20K, nearly all of these ($\sim 90\%$) are classified
as group members in \citet{Knobel2012}.

\begin{table*}
\caption{Wall volume groups catalog}
\label{tab:Tab4}  
\centering                  
\begin{tabular}{c c c c c c c c c c c c }    
\hline                
  ID & $Ra$ & $Dec$ & $z$ & ${\it N}_{obs}$ & ${\it N}_{corr}$ & $\sigma_{gap}$ & $L_{0.1-2.4keV}$ & $ M_{200}/M_{\sun}$ & $R_{200}$  & $ID_{Xray}$ & $Sep)$ \\   
              
     & (2000) & (2000) &   &               &                 & $(km\,s^{-1})$ & $(erg\,s^{-1})$ & $ $ & $ (Mpc)$  & & $ (\arcsec)$ \\   
\hline                    
   W1   &   149.9216  &   2.5175 &    0.7307 &  39  & 58.1 &  643 &  44.052 &  14.336 & 0.954 & COSMOS CL J095941.6+023129 & 7.9  \\ 
   W2   &   150.0684  &   2.4558 &    0.7304 &  20  & 31.5 &  322 &         &         &       &                            &      \\ 
   W3   &   150.0036  &   2.4509 &    0.7310 &  20  & 25.3 &  299 &  42.727 &  13.483 & 0.495 & COSMOS CL J100001.7+022712 & 17.0 \\
   W4   &   150.1422  &   2.0684 &    0.7266 &  17  & 23.1 &  361 &  43.010 &  13.667 & 0.571 & COSMOS CL J100035.2+020346 & 28.8 \\
   W5   &   150.1700  &   2.5230 &    0.6967 &  15  & 28.2 &  461 &  42.583 &  13.404 & 0.472 & COSMOS CL J100041.4+023124 & 14.5 \\
   W6   &   149.9807  &   2.5594 &    0.7316 &  14  & 25.2 &  323 &  42.611 &  13.41  & 0.469 & COSMOS CL J095956.6+023353 & 19.1 \\
   W7   &   149.9518  &   2.4819 &    0.7350 &  14  & 19.0 &  265 &  43.097 &  13.72  & 0.593 & COSMOS CL J095946.9+022908 & 17.9 \\
   W8   &   149.9178  &   2.5618 &    0.6999 &  11  & 24.7 &  316 &         &         &       &                            &      \\
   W9   &   150.0291  &   2.5625 &    0.7471 &  10  & 17.4 &  281 &  42.647 &  13.426 & 0.471 & COSMOS CL J100007.9+023308 & 36.6 \\
  W10   &   149.9890  &   2.5739 &    0.6956 &   9  & 17.9 &  238 &         &         &       &                            &      \\
  W11   &   150.0518  &   2.5983 &    0.6966 &   9  & 14.2 &  230 &         &         &       &                            &      \\
  W12   &   150.0294  &   2.3568 &    0.7261 &   9  & 18.0 &  386 &  42.489 &  13.332 & 0.442 & COSMOS CL J100005.8+022137 & 23.3 \\
  W13   &   149.8817  &   2.5764 &    0.6950 &   7  & 19.5 &  259 &  42.970 &  13.654 & 0.572 & COSMOS CL J095930.7+023440 & 7.9  \\
  W14   &   150.0885  &   2.2565 &    0.7280 &   7  &  9.3 &  303 &         &         &       &                            &      \\
  W15   &   150.0635  &   2.4133 &    0.7292 &   7  & 18.5 &  197 &         &         &       &                            &      \\
  W16   &   149.8979  &   2.5053 &    0.7350 &   7  & 12.5 &  139 &         &         &       &                            &      \\
  W17   &   150.0710  &   2.2823 &    0.7288 &   6  &  9.2 &  228 &         &         &       &                            &      \\
  W18   &   150.0394  &   2.2342 &    0.7251 &   5  &  6.7 &  177 &         &         &       &                            &      \\
  W19   &   150.1042  &   2.5524 &    0.7264 &   5  &  7.0 &  138 &         &         &       &                            &      \\
  W20   &   150.1068  &   2.5677 &    0.7292 &   5  &  9.7 &  138 &         &         &       &                            &      \\
  W21   &   150.1833  &   2.4233 &    0.7302 &   5  &  7.7 &  125 &         &         &       &                            &      \\
  W22   &   150.0361  &   2.4567 &    0.7319 &   5  &  8.1 &  189 &         &         &       &                            &      \\
  W23   &   149.7275  &   2.4802 &    0.7329 &   5  & 10.1 &  361 &         &         &       &                            &      \\
  W24   &   150.0534  &   2.2997 &    0.7331 &   5  &  7.9 &  339 &         &         &       &                            &      \\
  W25   &   149.7924  &   2.4895 &    0.7338 &   5  &  7.2 &  186 &         &         &       &                            &      \\
  W26   &   150.0017  &   2.2989 &    0.7523 &   5  &  5.7 &   64 &         &         &       &                            &      \\
\hline                             
\end{tabular}
\end{table*}

As a further check we matched our group list with the published list
of XMM-COSMOS extended sources presented in \citet{George2011}.  We
adopted as matching criterion a maximal angular distance on the sky of
$R_{200}$ in arcsecs, as estimated from X-ray data and listed in
\citet{George2011}, and a $\Delta z \sim 0.0035\times(1+z)$ in
redshift. Out of the nine groups in our catalog with ten or more
spectroscopic members, seven are associated with an X-ray detected group
in \citet{George2011}. Interestingly, the richest group we detect, which is group W1 in Table~\ref{tab:Tab4}, overlaps with the most luminous and
most massive X-ray extended source in \citet{George2011}.  Large
magenta circles on the left panel of Figure~\ref{Fig15} indicate the
sky position of these extended X-ray sources (size proportional to the
X-ray luminosity).

Vice versa, out of a total of 12 extended sources listed in
\citet{George2011} in the redshift range $0.69 \leq z_{spec} \leq
0.79$ and within our survey area, 9 possess a match with one of our
groups while 3 are unmatched.  Group W2 in Table~\ref{tab:Tab3},
however, is roughly $90\arcsec$ off one of these unmatched extended
X-ray sources, COSMOS CL J095908.0+022802, whose $R_{200}$ is $\sim
60\arcsec$. For this group we may have a centering problem due to our
incomplete sampling rate and/or overmerging of two distinct, but
adjacent, smaller structures.

For each of our groups possessing a match with those listed in
\citet{George2011}, we provide the corresponding X-ray information as
provided in \citet{George2011} in the last five columns of
Table~\ref{tab:Tab4}: the logarithm of the rest-frame luminosity in
the $0.1$-$2.4$ keV band; an estimate of the cluster mass,
log($M_{200}$), from weak-lensing-calibrated X-ray luminosity vs.
cluster mass scaling relation; an estimate of the cluster virial
radius, $R_{200}$ in Mpc; the \citet{George2011} cluster ID; the
distance in arcseconds between the centers of the two matched
clusters.

\section{Color and mass distribution of groups members}

As final sanity check on the quality of our group catalog, we searched
for the presence of the well-known interdependencies between
environment, galaxy masses, and colors \citep{Bolzonella2010,
  Cucciati2010, Iovino2010}. One of the final goals of our survey is
to provide a data set that enables a detailed exploration of the
interplay between the environment in which a galaxy resides and its
photometric and spectroscopic properties. We choose a selection as
close as possible to galaxy stellar mass of the sample observed to
enable an analysis down to yet unexplored low galaxy stellar
masses.

   \begin{figure}
    \centering
    \vspace{-1.4cm}
    \includegraphics[width=9.5cm, angle=0]{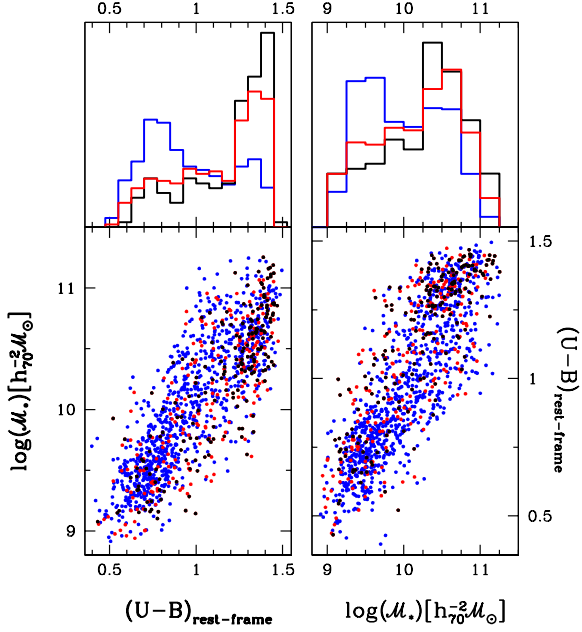}
    \vspace{-1.8cm}
    \caption{The bottom panels show the distribution in the plane
      galaxy stellar mass vs. $(U-B)$ rest-frame color (and vice
      versa) of all the galaxies of the Wall sample. Each galaxy is
      color coded according to the environment in which it resides:
      moving from blue to red and black, galaxies are located in
      progressively denser environments. The top panels show the
      normalized distribution in $(U-B)$ rest-frame colors (on the
      left) and in galaxy stellar masses (on the right) of these three
      samples.  Moving from lower to higher densities clearly results
      both in redder rest-frame color distributions and in stellar
      mass distributions skewed toward higher masses.}
       \label{Fig16}
     \end{figure}

While such a detailed study is beyond the scope of this paper, we
performed a simple analysis of the distribution of stellar masses and
$(U-B)$ rest-frame colors dividing the Wall total sample simply into
three broad categories. The first includes galaxies located in lower
density regions, which consist of either isolated galaxies that are not
in any of our detected groups or located in groups with corrected
richness $N_{corr} \leq 4$ and contain 892 galaxies; the second
includes galaxies in higher density regions, which encompass those located
in groups with corrected richness $ 4 < N_{corr} \leq 15$ and include 385
galaxies. Finally, we have a third sample including galaxies in the
highest density regions, defined as members of the richest
groups at $N_{corr} > 15$, with a total of 192 galaxies.

In Figure~\ref{Fig16} we show an outcome of this analysis.  The top
panels show the normalized distributions in $(U-B)$ rest-frame colors
(on the left) and in galaxy stellar masses (on the right) of these
samples. It is obvious that galaxies located in denser environments
progressively possess redder colors and higher masses, thus
reproducing the well-known trends that are found in the literature.
For a more detailed analysis of the spectrophotometric properties of
the COSMOS Wall galaxies as a function of their environment, we refer
to \citet{Petropoulou2016b}.

\section{Summary and conclusions}

We have presented the deep spectroscopic observations performed with
VIMOS at VLT of the large filamentary structure at z$\sim$0.73 in the
field of the COSMOS survey, also known as the COSMOS Wall. This structure
encompasses a comprehensive range of environments: from a dense
cluster and a number of galaxy groups to filaments, less dense regions,
and voids; thus this structure is of special interest for studies of environmental
effects on galaxy evolution.

We use photometric redshifts and the K-band selection to target
mass-selected galaxies with high probability of being within the
volume of interest. This strategy allows us to be very efficient (836
out of the 975 good redshift quality galaxies observed are in the
desired narrow redshift range $0.69\leq z \leq 0.79$) and to reach
completeness down to a stellar mass limit of log$({\cal M}_{*}/{\cal
  M}_{\odot})\sim 9.8$, significantly deeper than previously available
data. Thanks to a careful slit positioning strategy we obtained longer
exposure times for fainter targets, while retargeting, whenever
possible, brighter targets, thus ensuring good quality spectroscopic
data.

In this paper we detail the robust weighting scheme adopted to account
for the biases introduced by the photometric preselection of our
targets and we discuss our final galaxy catalog, together with its
physical properties, such as galaxy stellar mass and rest-frame magnitudes
estimates.

The final outcome of our survey is a sample of 1277 galaxies within
the COSMOS Wall volume, a number that doubles that available in this
region from previous surveys. As a result of the deeper galaxy stellar mass
limits we reached, we roughly quadruple the number of galaxies within
a mass complete sample.  In the \20K only $\sim 200$ out of 658
galaxies in the z-range $[0.69-0.79]$ form a complete sample down to
log$({\cal M}_{*}/{\cal M}_{\odot})\geq 10.7$, while in the COSMOS
Wall sample out of a total of 1277 galaxies, $\sim 800$ galaxies form
a complete sample down to log$({\cal M}_{*}/{\cal M}_{\odot})\geq
9.8$.

We performed a simple sanity check, for both our weighting scheme and
our stellar mass estimates, by computing the number density of
galaxies in bins of galaxy stellar mass and comparing this number with results
available in the literature. The agreement is very good: while the
region of the Wall structure displays a significant overdensity as
expected, when this region is excised from the analysis the number
densities as a function of galaxy stellar mass agree well with those
in the literature.

Our new sample enables us to perform a detailed, high definition,
mapping of the complex Wall structure, and we have built a new group
catalog using a FOF algorithm and a multipass procedure.  The
agreement between the largest groups we detect and the presence of
extended X-ray emission from deep published catalogs is remarkable.  A
simple preliminary exploration of galaxy colors and masses as a
function of the environment defined using this groups catalog,
reassuringly displays the trends well known in the literature.

The accurate environmental information coupled with good quality
spectroscopic information and the rich ancillary data available in the
COSMOS field is therefore a gold mine for detailed studies of galaxy
properties as a function of local environment in a redshift slice
where environmental effects have been shown to be important, and in
the mass range where mass and environment driven effects are both at
work.

The COSMOS Wall sample presented in this paper thus provides a
valuable laboratory for the accurate mapping of environmental effects
on galaxy evolution at a look-back time of $\sim$6.5 Gyr, when the
Universe was roughly half its present age. 

\begin{acknowledgements}

This project was made possible by INAF funding through the PRIN 2011
program. The authors wish to acknowledge generous support by ESO
staff during service observations. We warmly thank Chris Haines for his
useful suggestions and comments.  

\end{acknowledgements}


\bibliographystyle{aa}
\bibliography{AIovino}

\end{document}